\documentclass[aps,pre,twocolumn,showpacs,eqsecnum]{revtex4}
\usepackage{amssymb}
\usepackage{amsmath}

\newcommand{\bq}{\begin{equation}}
\newcommand{\nq}{\end{equation}}
\input{tcilatex}

\begin{document}

\title{Anomalous relaxation in quantum systems and the non-Markovian
stochastic Liouville equation.}
\author{A. I. Shushin}
\affiliation{Institute of Chemical Physics, Russian Academy of Sciences, 117977, GSP-1,
Kosygin str. 4, Moscow, Russia}

\begin{abstract}
The kinetics of relaxation in quantum systems induced by anomalously slowly
fluctuating noise is studied in detail. In this study two processes are
considered, as examples: (1) relaxation in two-level system (TLS) caused by
external noise with slowly decaying correlation function $P (t)
\sim (wt)^{-\alpha}$, where $0 <\alpha < 1$, and (2) anomalous-diffusion
controlled radical pair (RP) recombination in which relaxation results from
the reaction with slowly fluctuating reaction rate whose
fluctuations are governed by subdiffusive relative motion of radicals in a
potential well. Analysis of these two processes is made within continuous
time random walk approach (CTRWA). Rigorous CTRWA-treatment of the processes
under study results in the non-Markovian stochastic Liouville equation (SLE)
for the density matrix of the systems. This SLE predicts important specific
features of relaxation kinetics of quantum systems in the presence of the
above mentioned anomalous noise. In TLS, for example, relaxation of both
phase and population turns out to be anomalously slow and strongly
non-exponential. Moreover, for $\alpha < 1$ in the limit of characteristic
fluctuation rates $w$ much larger than the frequency of quantum transitions $%
\omega_s$ ($w/\omega_s \gg 1$) the relaxation kinetics in TLS is independent
of $w$. Strong changes of the relaxation kinetics is found with the change
of $w$ and $\alpha$. In RP recombination the anomalous fluctuations of
reaction rate (due to anomalous diffusion) show themselves in non-analytical
dependence of the reaction yield $Y_r$ on reactivity and parameters of the
RP spin Hamiltonian. In particular, the spectrum shape of reaction yield
detected magnetic resonance, i.e. the dependence of $Y_r$ on the frequency $%
\omega$ of resonance microwave field is found to be strongly non-Lorenzian
with the width to be determined by the strength $\omega_1$ of this field.
\end{abstract}

\pacs{PACS numbers: 05.40.Fb, 02.50.-r, 76.20.+q}
\maketitle

\bigskip

\section{Introduction}

The noise induced relaxation in quantum systems is very important process,
which is investigated in various brunches of physics and chemistry: magnetic
resonance \cite{Abr}, quantum optics, nonlinear spectroscopy \cite{Muk},
solid state physics \cite{Hau,Bou}, etc.

Some of these processes are analyzed assuming fast decay of correlation
functions of this noise $\tau_c$ and considering the short correlation time
limit (SCTL). This approach is well known in the general relaxation theory
\cite{Gard}. In the absence of memory the relaxation is described by very
popular Bloch-type equations. Nevertheless, in a large number of other
processes the memory effects come to play. Important steps beyond the
conventional short correlation time approximation can be made within the
approach treating the relaxation kinetics with the stochastic Liouville
equation (SLE) \cite{Kubo}. This approach, however, also has strong
limitations: it assumes the noise fluctuations to be Markovian. The
alternative approach, which can be applied for the analysis memory effects
is based on the Zwanzig projection operator formalism \cite{Fors}.
Unfortunately, in reality the projection formalism allows for tractable
analysis of the memory effects only in the lowest orders in the fluctuating
interaction $V$ inducing relaxation \cite{Argy}.

At present especially significant interest to the correct interpretation of
long memory (non-Markovian) effects has been excited by recent works
concerning processes governed by noises whose correlation functions $P (t)$
decay anomalously slowly: $P (t) \sim t^{-\alpha}$ with $\alpha < 1$. A
large number of different phenomena, in which these processes play important
role, are considered in the review article \cite{Met}. Some of them are
discussed in relation to spectroscopic studies of quantum dots \cite%
{Shim,Silb}. Similar problems are analyzed in the theory of stochastic
resonances \cite{Han}. Many works study the manifestation of non-Markovian
long memory effects in dielectric relaxation and linear dielectric response
\cite{Tit} (and references therein). All anomalous relaxation processes
mentioned above cannot be properly described by any methods based on the
short correlation time approximation. The conventional SLE approach is not
appropriate for description of these processes either.

The efficient method of analyzing the memory effects (including the case of
anomalously long memory) has recently been developed in ref. \cite{Shu1}. It
treats the noise within the continuous time random walk approach (CTRWA)
\cite{Scher1,Scher2,Wei}, describing the statistical properties of the noise
in terms of the probability distribution functions (PDFs) of renewals of the
interaction fluctuations associated with the noise. The developed
CTRWA-based method is based on the Markovian representation of the any
CTRW-processes by the Markovian ones proposed in ref. \cite{Shu1}. This
representation allows for obtaining the formal matrix expression for the
evolution operator of the system under study, which can be considered as the
non-Markovian generalization of the conventional SLE \cite{Shu1} and is
called hereafter the non-Markovian SLE.

The non-Markovian SLE appeared to be very fruitful for description of some
classical processes assisted by stochastic anomalous spatial migration \cite%
{Shu2}. The memory effects are known to manifest themselves in this
processes extremely strongly leading to strongly non-exponential anomalous
relaxation kinetics \cite{Blum,Sung,Tach}.

In this work the proposed non-Markovian SLE is applied to the detailed
analysis of specific features of relaxation in quantum systems induced by
anomalous noise, i.e. the noise with anomalously long-tailed correlations in
time, whose effect on quantum systems can hardly be correctly treated by
expansion in powers of the noise amplitude. This type of the SLE enables one
to describe relaxation kinetics without expansion in the fluctuating
interaction. In some physically reasonable approaches, for example in the
sudden relaxation model (SRM), it allows for describing the phase and
population relaxation kinetics in analytical form. In the work two types of
processes are considered: relaxation in two-level systems (TLS) and
anomalous-diffusion assisted radical pair (RP) recombination in a potential
well, as an example of processes in multilevel systems. Relaxation in these
systems is shown to be strongly non-exponential. In addition, in RP
recombination anomalous properties of relaxation (caused by anomalous
relative diffusion) result in some peculiarities of magnetic field effects
\cite{St}: non-analytical dependence of magnetic field affected
recombination yield (MARY) on the parameters of the spin Hamiltonian,
strongly non-Lorenzian shape of lines of reaction yield detected magnetic
resonance (RYDMR) \cite{St}, etc.

\section{General formulation}

We consider noise induced relaxation in the quantum system whose evolution
is governed by the hamiltonian
\begin{equation}
H (t) = H_s + V (t),  \label{gen1}
\end{equation}
where $H_s$ is the term independent of time and $V (t)$ is the fluctuating
interaction, which models the noise. The evolution is described by the
density matrix $\rho (t)$ satisfying the Liouville equation ($\hbar = 1$)
\begin{equation}
\dot \rho = - i {\hat H} (t) \rho, \;\: \mbox{with} \;\: {\hat H} \rho = [
H,\rho ] = [ H \rho - \rho H ].  \label{gen2}
\end{equation}

${V} (t)$-fluctuations are assumed to be symmetric ($\langle V \rangle = 0$)
and result from stochastic jumps between the states $|x_{\nu} \rangle$ in
the (discrete or continuum) space $\{ x_{\nu} \} \equiv \{ x \}$ with
different $V = V_{\nu}$ and $H = H_{\nu}$ (i.e. different $\hat V = \hat
V_{\nu} \equiv [V_{\nu}, \dots] $ and $\hat H = \hat H_{\nu}$):
\begin{equation}
\hat {\mathcal{V}} = \sum\nolimits_{\nu} \! |x_{\nu} \rangle \hat V_{\nu}
\langle x_{\nu} | \;\,\mbox{and} \;\, \hat {\mathcal{H}} =
\sum\nolimits_{\nu} \! |x_{\nu} \rangle {\hat H}_{\nu} \langle x_{\nu} |.
\label{gen4}
\end{equation}

Hereafter we will apply the bra-ket notation:
\begin{equation}
|k\rangle, \,\;\;|k k^{\prime}\rangle \equiv |k\rangle \langle k^{\prime}|,
\;\;\mbox{and} \;\; |x\rangle  \label{gen4a}
\end{equation}
for the eigenstates of $H$ (in the original space) and ${\hat H}$ (in the
Liouville space), and for states in $\{ x \}$-space, respectively.

It is worth noting that within the proposed semiclassical approach one can
also describe some fluctuating irreversible processes in the system modelled
by the additional non-Hermitian relaxation term $-i\hat {K} (t)$ in the
Lioville operator $\hat { H}$. In this model the effect of fluctuating
irreversible process results in modification of the Liouville equation (\ref%
{gen2}):
\begin{equation}
\dot \rho = - i \hat L \rho, \;\;\mbox{in which} \;\; \hat L = \hat {%
\mathcal{H}} -i \hat {\mathcal{K}}.  \label{gen4b}
\end{equation}

In the majority of processes the time evolution of observables under study
is determined by the evolution operator ${\hat {\mathcal{R}}} (t)$ in the
Liouville space averaged over $V (t)$- and $K (t)$-fluctuations:
\begin{equation}
\rho (t) = {\hat {\mathcal{R}}} (t) \rho_i, \,\; \mbox{where} \;\;{\hat {%
\mathcal{R}}} (t) = \left\langle T e^{-i \! \int_0^{t} \!d\tau \hat {L}
(\tau)} \right\rangle.  \label{gen5a}
\end{equation}
The operator ${\hat {\mathcal{R}}} (t)$ can equivalently be represented in
terms of the conditional evolution operator $\hat {\mathcal{G}}
(x,x^{\prime}|t)$ averaged over the initial distribution $P_i (x)$:
\begin{equation}
{\hat {\mathcal{R}}} (t) = \langle \hat {\mathcal{G}} \rangle \equiv
\!\sum\nolimits_{x, x_i} \!\!\hat {\mathcal{G}} (x,x_i|t) P_i (x_i).
\label{gen5b}
\end{equation}
For steady state $V(t)$-fluctuations the averaging should be made over the
equilibrium distribution $P_e (x)$, i.e. $P_i (x) = P_e (x)$.

It is worth noting, however, that evaluation of some observables requires
analysis of the conditional operator $\hat {\mathcal{G}} (x,x_i|t)$ itself
rather than the averaged one ${\hat {\mathcal{R}}} (t)$ (see Sec. IVC).

Thus, in the proposed semiclassical approximation for $V(t)$- and $K (t)$%
-fluctuations the relaxation kinetics is determined by the operator $\hat {%
\mathcal{G}} (x,x_i|t)$. Its evaluation is, in general, a complicated
problem. In what follows we will discuss important approaches in which this
problem can be significantly simplified by reducing it to solving the
(differential or integral) SLE for $\hat {\mathcal{G}} (x,x_i|t)$ \cite{Kubo}%
.

In our further study we will often use the Laplace transformation of
functions under study in time $t$ conventionally denoted as
\begin{equation}
{\widetilde Z}(\epsilon) = \int_0^{\infty} \! dt \, Z(t)e^{-\epsilon t}
\label{gen6}
\end{equation}
for any function $Z(t)$.

\section{Stochastic Liouville equations}

The goal of our work is to analyze the strong memory effects within the
CTRWA. This approach is known to result in the complicated non-Markovian SLE
\cite{Shu1} for $\hat {\mathcal{G}} (x,x_i|t)$. However, it is instructive
to start the discussion with the simpler Markovian models reducing to the
conventional semiclassical SLE.

\subsection{Markovian fluctuations}

In the Markovian approach the kinetics of jumps in $\{ x \}$-space leading
to $V(t)$-fluctuations are described by the PDF $P (x,t|x_i,t_i)$ satisfying
equation \cite{Gard}
\begin{equation}
\dot P = - \hat {\mathcal{L}} P \;\;\; \mbox{with} \;\;\; P (x,t_i|x_i,t_i)
= \delta_{x x_i},  \label{mark1}
\end{equation}
where $\hat {\mathcal{L}}$ is some linear operator. The principal
simplification of the problem results from the fact that in the Markovian
approach (\ref{mark1}) the ${\hat {\mathcal{G}}} (x,t|x_i,t_i)$ obeys the
SLE \cite{Kubo}:
\begin{equation}
\dot {\hat {\mathcal{G}}} = -(\hat {\mathcal{L}}+i\hat {L}) {\hat {\mathcal{G%
}}}, \;\:\mbox{where} \;\;{\hat {\mathcal{G}}} (x,t_i|x_i,t_i) = \delta_{x
x_i},  \label{mark2}
\end{equation}
so that in accordance with eq. (\ref{gen5b}) we get for the Laplace
transform $\hat {\widetilde {\mathcal{R}}}$:
\begin{equation}
\hat {\widetilde {\mathcal{R }}} = \langle \hat {\widetilde {\mathcal{G}}}
\rangle = \langle (\epsilon + i\hat {L} + \hat {\mathcal{L}})^{-1} \rangle
\label{mark2a}
\end{equation}

The idea of the SLE (\ref{mark2}) is based on the simple observation that
the changes of $\hat {\mathcal{G}} $ resulted from dynamical motion [eq. (%
\ref{gen2})] and stochastic evolution [eq. (\ref{mark1})] during short time $%
\Delta t$ are written as
\begin{equation}
\Delta_d \hat {\mathcal{G}} = - (i\hat {L}\hat {\mathcal{G}}) \Delta t\;\;%
\mbox{and} \;\; \Delta_f \hat {\mathcal{G}} = - (\hat {\mathcal{L}})\hat {%
\mathcal{G}} )\Delta t,  \label{mark3}
\end{equation}
respectively, so that total change of $\hat {\mathcal{G}}$ is given by the
expression
\begin{equation}
\Delta \hat {\mathcal{G}} = \Delta_f \hat {\mathcal{G}} + \Delta_d \hat {%
\mathcal{G}} = -(\hat {\mathcal{L}} + i\hat {L })) \hat {\mathcal{G}} \Delta
t,  \label{mark4}
\end{equation}
equivalent to eq. (\ref{mark2}).

\subsection{Non-Markovian fluctuations}

\subsubsection{Continuous time random walk approach}

Non-Markovian $V (t)$-fluctuations can conveniently be described by the
CTRWA [which leads to the non-Markovian SLE \cite{Shu1} for $\hat {\mathcal{G%
}} (t)$]. It treats fluctuations as a sequence of sudden changes of $\hat V $%
. The onset of any particular change of number $j$ is described by the
matrix $\hat P_{j-1}$ (in $\{x\}$-space) of probabilities not to have any
change during time $t$ and its derivative $\hat W_{j-1} (t) = - d\hat
P_{j-1} (t)/ dt$, i.e. the PDF-matrix for times of waiting for the change.
These matrices are diagonal and independent of $j$ at $j > 1$:
\begin{equation}
\hat P_{j-1} (t) = \hat P (t), \: \hat W_{j-1} (t) = \hat W (t) = - d\hat P
(t)/dt.  \label{nmark1}
\end{equation}
For $j = 1$
\begin{equation}
\hat P_{0} (t) \equiv \hat P_i (t)\;\;\mbox{and} \;\;\hat W_{0} (t) \equiv
\hat W_{i} (t) = - d\hat P_{i} (t)/dt  \label{nmark2}
\end{equation}
depend on the problem considered. For non-stationary ($n$) and stationary ($%
s $) fluctuations \cite{Scher1,Scher2,Wei}
\begin{eqnarray}
\hat W_i (t) &=& \hat W_n (t) = \hat W (t), \\
\hat W_i (t) &=& \hat W_s (t) = {\hat \tau_e}^{-1}
\mbox{$\int_t^{\infty} \!
d\tau \, \hat W (\tau)$},
\end{eqnarray}
respectively, where $\hat \tau_e = \int_0^{\infty} \! dt \, t \hat W (t)$ is
the matrix of average times of waiting for the change \cite%
{Scher1,Scher2,Wei}.

It is worth noting some relations for the Laplace transforms of $\hat W_j
(t) $ and $\hat P_j (t)$ suitable for our further analysis: $\hat {%
\widetilde P}_j (\epsilon)= [1 - \hat {\widetilde W}_j (\epsilon)]/\epsilon$
and $\hat {\widetilde W}_s (\epsilon)= \hat {\widetilde P} (\epsilon)/\hat
\tau$ \cite{Scher1,Scher2,Wei}, as well as the representations
\begin{equation}
\hat {\widetilde {W}} (\epsilon) = [1 + \hat \Phi (\epsilon)]^{-1} \;\: %
\mbox{and} \;\: \hat {\widetilde P} (\epsilon) = [\epsilon + \epsilon/\hat
\Phi (\epsilon)]^{-1}.  \label{nmark4}
\end{equation}
in terms of the auxiliary matrix $\hat \Phi (\epsilon)$ diagonal in $\{ x\}$%
-space (see below).

\subsubsection{Markovian representation of CTRWA}

In this Section we will briefly discuss the Markovian representation of the
CTRWA recently proposed in ref. \cite{Shu1}, which appears to be very useful
for the CTRWA-based analysis of the problem under study and, in particular,
provides the most rigorous method of deriving the non-Markovian SLE.

Suppose that the kinetics of $(\nu \to \nu^{\prime})$-transitions in $\{x\} $%
-space is controlled by the Markovian process in another $\{q_j\} $-space,
which is governed by the operator $\hat \Lambda$. The corresponding PDF $%
\sigma (j,t)$ satisfies equation
\begin{equation}
\dot \sigma = - \hat \Lambda \sigma  \label{ma1}
\end{equation}
describing evolution in $\{q_j\}$-space and equilibration if the operator $%
\hat \Lambda$ has the equilibrium state $|e_q\rangle$ ($\hat \Lambda
|e_q\rangle = 0$). This state is represented as
\begin{equation}
|e_q\rangle = \sum\nolimits_{j} p_{q_{j}}^e |j\rangle, \;\langle e_q | =
\sum\nolimits_{j} \langle j |, \; (\langle e_q | e_g\rangle = 1).
\label{ma1a}
\end{equation}
where $p_j^e$ are the probabilities of equilibrium population of states in $%
\{q_j\} $-space.

The $q_j$-process is assumed to control the evolution in $\{x\} $-space as
follows: $(\nu \to \nu^{\prime})$-transitions occur with the rate $%
\kappa_{\nu^{\prime}\nu}$ whenever the system visits the transition state $|
t \rangle$ in $\{q_j\}$-space. Transitions can lead to the change in $| j
\rangle$-state, i.e to $(|t \rangle \to | n \rangle)-$transition with $|n
\rangle \neq | t \rangle$. For simplicity, we assume that $|n \rangle = | t
\rangle$ as well as that $\hat \Lambda$ and $| t \rangle$-state are
independent of the state in $\{x\}$-space.

The evolution of the system in $\{ x \otimes q_j\}-$space is described by
the PDF matrix $| \hat \rho \rangle$ obeying the SLE
\begin{equation}
|\dot {\hat \rho} \rangle = -({\hat \Lambda} + i\hat {L} + {\hat K_d} - {%
\hat K_o}) |\hat \rho \rangle ,  \label{ma2}
\end{equation}
where
\begin{equation}
{\hat K_d} = \hat \kappa_d \otimes |t\rangle \langle t | \; \mbox{and} \; {%
\hat K_o} = \hat \kappa_o \otimes |n \rangle \langle t |  \label{ma3}
\end{equation}
are the transition matrices (operating in $\{ x\otimes q_j\}$-space) are
diagonal and non-diagonal in $\{x\}$-subspace, respectively, in which
\begin{equation}
\hat \kappa_d = \sum\nolimits_{\nu} \!|\nu\rangle \kappa_{\nu\nu}\langle \nu
|, \;\;\hat \kappa_o = \sum\nolimits_{\nu, \nu^{\prime}\neq \nu}
\!|\nu\rangle \kappa_{\nu\nu^{\prime}}\langle \nu^{\prime}| .  \label{ma5}
\end{equation}
and $\kappa_{\nu\nu} = \sum\nolimits_{\nu^{\prime}(\neq \nu)}
\kappa_{\nu^{\prime}\nu}$. Equation (\ref{ma2}) should be solved with the
initial condition
\begin{equation}
|\hat \rho\rangle_{t=0} = | i \rangle \sum\nolimits_{\nu}|\nu \rangle
\langle \nu|,  \label{ma6a}
\end{equation}
where $| i \rangle = \sum\nolimits_j p_{q_j}^{i} |j\rangle, \;\,\mbox{with}%
\;\,\langle e_q | i \rangle = \sum\nolimits_j p_{q_j}^{i} = 1$.

The function of interest is the (time) Laplace transformed PDF in $\{ x \}$%
-space
\begin{equation}
\hat {\widetilde {\mathcal{G}}} = \langle e_q | \hat {\widetilde \rho}
\rangle = \sum\nolimits_j \hat {\widetilde \rho}_j, \;\;\mbox{in which}\;\;
\hat {\widetilde \rho}_j = \langle j | \hat {\widetilde \rho} \rangle.
\label{ma6b}
\end{equation}
It is determined by solution $|\hat {\widetilde{\rho}} (\epsilon)\rangle$ of
equation
\begin{equation}
|\hat {\widetilde \rho} \rangle = \hat G |i\rangle + \hat G \hat K_o |\hat {%
\widetilde \rho} \rangle,  \label{ma7}
\end{equation}
where
\begin{equation}
\hat G = (\hat \Omega + \hat \Lambda + \hat K_d)^{-1} \;\;\mbox{with}\;\;
\hat \Omega = \epsilon + i\hat {L}.  \label{ma8}
\end{equation}

The expression for $\hat {\widetilde {\mathcal{G}}}$, obtained by solving
eq. (\ref{ma7}), can be represented in a CTRWA-like form, which in what
follows [by analogy with the Markovian model (\ref{mark1})-(\ref{mark2a})]
will be treated as the solution of the so called non-Markovian SLE \cite%
{Shu1}.

Hereafter, for brevity, we will sometimes omit the argument $\hat \Omega$ of
the Laplace transforms of functions under study if this does not result in
confusions.

\subsubsection{Non-Markovian SLE}

Solution of eq. (\ref{ma7}) leads to the following non-Markovian SLE,
written in matrix resolvent form \cite{Shu1},
\begin{equation}
\hat {\widetilde {\mathcal{G}}} = \hat {\widetilde P}_i (\hat \Omega) +
\hat {\widetilde P}_n (\hat \Omega) [1 - \hat{\mathcal{P}}\hat {\widetilde W}%
_n (\hat \Omega) ]^{-1} \hat {\mathcal{P}}\hat {\widetilde W}_{i} (\hat
\Omega),  \label{ma9}
\end{equation}
where
\begin{equation}
\hat{\mathcal{P}} = \hat \kappa_o/\hat\kappa_d \;\;\mbox{with}
\;\;\sum\nolimits_{\nu} \mathcal{P}_{\nu\nu^{\prime}} = 1  \label{ma10}
\end{equation}
is the matrix of jump probabilities with zeroth diagonal elements and $\hat {%
\widetilde W}_{\mu}\; (\mu = n, i )\;$ are written as
\begin{equation}
\hat {\widetilde W}_{\mu} = \hat \kappa_d \hat G_{\mu}, \; \mbox{with} \;
\hat G_{\mu} = (1+\hat g_t \hat \kappa_d)^{-1} \hat g_{\mu}  \label{ma16a}
\end{equation}
in which
\begin{equation}
\hat g_{\mu} = \langle t | (\hat \Omega + \hat \Lambda)^{-1} | \mu \rangle,
\;\;\, \hat g_{t} = \langle t | (\hat \Omega + \hat \Lambda)^{-1} |t \rangle.
\label{ma16c}
\end{equation}

The non-Markovian SLE (\ref{ma9}) can also be represented in an alternative
convenient form \cite{Shu1}
\begin{eqnarray}
\hat {\widetilde {\mathcal{G}}} &=& \hat {\widetilde P}_i + \hat \Omega^{-1}
\hat \Phi (\hat \Phi + \hat {\mathcal{L}})^{-1} \hat {\mathcal{P}} \hat {%
\widetilde W}_{i} \\
&=& \hat {\widetilde P}_{m_i} + \hat \Omega^{-1} \hat \Phi (\hat \Phi +
\hat {\mathcal{L}})^{-1} \hat {\widetilde W}_{m_i},  \label{ma17}
\end{eqnarray}
where
\begin{equation}
\hat {\mathcal{L}} = 1 - \hat {\mathcal{P}} = 1 - \hat \kappa_o/\hat\kappa_d
\label{ma17a}
\end{equation}
is the Kolmogorov-Feller-type operator describing the jump-like motion in $%
\{x\}$-space,
\begin{equation}
\hat \Phi = (\hat g_t-\hat g_n)\hat g_n^{-1} + (\hat \kappa_d \hat g_n)^{-1}
\label{ma17b}
\end{equation}
is the matrix function (diagonal in $\{x\}$-space) characterizing $\hat W
(t) $ [see eq. (\ref{nmark4})]: $\hat {\widetilde W} = \hat {\widetilde W}_n
= (1+\hat \Phi)^{-1}$, and
\begin{equation}
\hat {\widetilde W}_{m_i} = \hat {\widetilde W}_{i}/\hat {\widetilde W}\;\;%
\mbox{and}\;\; \hat {\widetilde P}_{m_i} = (1 - \hat {\widetilde W}%
_{m_i})/\hat \Omega  \label{ma17bb}
\end{equation}
are auxiliary (modified) PDF matrices of the type of waiting time PDFs.

The representation (\ref{ma17}) of the non-Markovian SLE is somewhat
different from that proposed in ref. \cite{Shu1}. Both representation are,
however, equivalent and can equally be used for the analysis, though eq. (%
\ref{ma17}) is closer in its form to the CTRWA-like expressions \cite{Hau}.

The operator $\hat {\mathcal{L}}$ is assumed to have the equilibrium
eigenstate $|e_x^0 \rangle$ (i.e. $\hat {\mathcal{L}} |e_x^0 \rangle = 0$):
\begin{equation}
|e_x^0 \rangle = \sum\nolimits_x P_e^0 (x)|x \rangle \;\;\mbox{ and }
\;\;\langle e_x^0 | = \sum\nolimits_x \langle x |.  \label{ma17c}
\end{equation}
The eigenstate $|e_x^0 \rangle$, however, is not the true equilibrium state
of the system under study. The true equilibrium state $|e_x \rangle$ is also
determined by the behavior of $\hat \Phi (\epsilon)$ at $\epsilon \to 0$. It
is clear from eq. (\ref{ma17b}) and definition of the PDF matrix $\hat W (t)$
that $\hat \Phi (0) = 0$, therefore, in general, we can write
\begin{equation}
\hat \Phi (\epsilon) \overset{\epsilon \to 0}{\sim} (\epsilon/\hat w)^{\hat
\alpha}, \;\;\mbox{where}\;\; \hat w = \mbox{$\sum\nolimits_x$} |x \rangle
w_x \langle x |.  \label{ma17d}
\end{equation}
In this equation the matrices $\hat \alpha $ (of exponents) and $\hat w$ [of
characteristic rates (see Sec. V)] are assumed to be diagonal in $\{ x \}$%
-space. Moreover, for simplicity, to avoid analysis of exotic equilibrium
states \cite{Shu1} we assume that $\hat \alpha \equiv \alpha$ is just a
parameter independent of $x$ rather than matrix. In accordance with formulas
(\ref{ma17})-(\ref{ma17b}), for such $\alpha\,$ $\hat {\widetilde {\mathcal{G%
}}} \overset{\epsilon \to 0}{\sim} (\epsilon + \hat {\mathcal{L}}{\hat w}%
^{\alpha}\epsilon^{1-\alpha})^{-1}$, i.e. the state $|e_x \rangle$ is,
actually, the equilibrium eigenstate of $\hat {\mathcal{L}}{\hat w}^{\alpha}$
($\hat {\mathcal{L}}\hat w^{\alpha}|e_x \rangle = 0$), which is written as
\begin{equation}
|e_x \rangle = N_w^{-1} \hat w^{\!-\alpha}|e_x^0 \rangle \;\;\mbox{with}
\;\; N_w = \langle e_x^0 |\hat w^{\!-\alpha}|e_x^0 \rangle.  \label{ma17e}
\end{equation}

Possible expressions for $|e_x \rangle$ in some particular models of $\hat {%
\mathcal{L}}$ are discussed below in Sec. IVA.

It is important to note that with the eigenstate $|e_x \rangle$ the average
of any operator $\hat Y$ can be represented as
\begin{equation}
\langle \hat Y \rangle = \langle e_x| \hat Y |e_x\rangle.  \label{ma17e1}
\end{equation}
In particular, as it follows from eq. (\ref{gen5a}),
\begin{equation}
{\hat {\mathcal{R}}} (t) = \langle e_x| \hat {\mathcal{G}} |e_x\rangle
\equiv \langle \hat {\mathcal{G}} \rangle.  \label{ma17f}
\end{equation}

According to eqs. (\ref{ma9})-(\ref{ma16a}) the initial state $|i\rangle$
(in $\{ q_j \}$) manifests itself only in the expressions for matrices $\hat
W_i (t)$ and $\hat P_i (t)$. In particular \cite{Scher1,Scher2,Wei}:

a) In the non-stationary $n$-CTRWA $|i \rangle = |n \rangle$, so that $\hat {%
\widetilde W}_{i} = \hat {\widetilde W}_{n}$, $\hat {\widetilde P}_{i} =
\hat {\widetilde P}_{n}$ and
\begin{equation}
\hat {\widetilde {\mathcal{G}}} (\hat \Omega) = \hat {\widetilde {\mathcal{G}%
}}_n (\hat \Omega)= \hat \Omega^{-1}\hat \Phi (\hat\Omega) [\hat \Phi
(\hat\Omega) + \hat {\mathcal{L}}]^{-1}.  \label{ma18}
\end{equation}

b) In the stationary $s$-CTRWA $|i \rangle = |e \rangle$, where $| e \rangle
$ is the equilibrium eigenstate [see eq. (\ref{ma1a})], consequently $\hat {%
\widetilde W}_i = \hat {\widetilde W}_s = \hat {\widetilde P}_n /\hat \tau,
\;$ where $\; \hat \tau = \hat g_n \hat \Phi /p_t^e \;$ is the matrix of
average times (diagonal in $\{ x \}$-space) with $\: p_t^e = \langle t | e
\rangle$. In the considered case $|n \rangle = |t\rangle$, when $\hat \Phi =
1/(\hat \kappa_d \hat g_t)$ and $\hat \tau = 1/(\hat \kappa_d p_t^e )$.
Substitution of $\hat {\widetilde W}_s$ into eq. (\ref{ma9}) yields
\begin{equation}
\hat {\widetilde {\mathcal{G}}} (\hat \Omega) = \hat {\widetilde {\mathcal{G}%
}}_s (\hat \Omega) = \hat \Omega^{-1} - \hat {\widetilde {\mathcal{G}}}_n
(\hat \Omega) {\hat {\mathcal{L}}}(\hat \Omega \hat \tau)^{-1},
\label{ma19a}
\end{equation}

\section{Useful models and approaches}

\subsection{Models for jump motion}

\subsubsection{Sudden relaxation model (SRM).}

The SRM \cite{Shu1} assumes sudden equilibration in $\{x\}$-space described
by operator
\begin{equation}
\hat {\mathcal{L}} = (1 - |e_0 \rangle \langle e_0 |)\hat Q^{-1},\;\; \hat Q
= 1-\mbox{$\sum\nolimits_x$} P_x |x \rangle \langle x |,  \label{sud1}
\end{equation}
in which
\begin{equation}
|e_0 \rangle = \mbox{$\sum\nolimits_x\!  P_x |x \rangle$},\:\;\mbox{and}\;\:
\langle e_0 |=\mbox{$\sum\nolimits_x \!\langle x |$}.  \label{sud1a}
\end{equation}
is some auxiliary vector determined by the equilibrium vector $|e_x \rangle$%
:
\begin{equation}
|e_x \rangle = \mbox{$\sum\nolimits_x  P_e (x) |x \rangle$} = \hat q |e_0
\rangle \,\; \mbox{and}\;\,\langle e_x |=\langle e_0 |,  \label{sud2}
\end{equation}
where
\begin{equation}
\hat q = N_0^{-1}\hat Q \hat w^{-\alpha} \;\; \mbox{with} \;\; N_0 = %
\mbox{$\sum\nolimits_x [(P_x - P_x^2)/w_x^{\alpha}]$}.  \label{sud2a}
\end{equation}

According to eqs. (\ref{sud2}) and (\ref{sud2a}) :
\begin{eqnarray}
P_x &=& \mbox{$\frac{1}{2}$} - \sqrt{\mbox{$\frac{1}{4}$}-N_0 w_x^{\alpha}
P_e(x)}.  \label{sud2b2}
\end{eqnarray}
This relation determines the vector $|e_0 \rangle$ (in the definition of $%
\hat {\mathcal{L}}$) which ensures the given equilibrium state $|e_x \rangle$%
. The value of the parameter $N_0$ is fixed by the normalization condition
for the distribution $P_x$:
\begin{equation}
\sum\nolimits_x\Big[\mbox{$\frac{1}{2}$} - \sqrt{\mbox{$\frac{1}{4}$}-N_0
w_x^{\alpha} P_e (x)}\Big] = 1.  \label{sud2c}
\end{equation}

In the model (\ref{sud1}) one gets for any $\hat {\widetilde{W}}_i$
\begin{equation}
\hat {\widetilde{\mathcal{R}}}_i = \langle \hat {\widetilde{P}}_{\!Q_i}
\rangle + \langle {\hat q^{-1}\widetilde{P}}_{\!Q} \rangle [1 - \langle \hat
q^{-1} \hat {\widetilde{W}}_{\!Q} \rangle]^{-1} \langle \hat {\widetilde{W}}%
_{\!Q_i} \rangle,  \label{sud3}
\end{equation}
where $\hat {\widetilde{P}}_{\!Q_i} = (1 - \hat {\widetilde{W}}%
_{\!Q_i})/\hat \Omega$,
\begin{equation}
\hat {\widetilde{W}}_{\!Q} = (1+ \hat \Phi \hat Q)^{-1}, \;\; \hat {%
\widetilde{W}}_{\!Q_i} = \hat {\widetilde{W}}_i (\hat {\widetilde{W}}_{\!Q}/%
\hat {\widetilde{W}}).  \label{sud4}
\end{equation}

The obtained general formulas are simplified in some particular models, for
example, in the $N$-state SRM with
\begin{equation}
|e_0 \rangle = N^{-1} \mbox{$\sum\nolimits_x \! |x \rangle$}, \;\hat Q =
Q_{\!N} = 1-N^{-1},  \label{sud5}
\end{equation}
In this model $\hat q = \hat w^{\!-\alpha}/\sum\nolimits_x w_x^{\!-\alpha}$.
It is important to note that in the case of two states the model (\ref{sud5}%
) is the only possible one.

Particularly simple expressions are obtained in the model of the only
characteristic fluctuation rate:
\begin{equation}
\hat w \equiv w, \;\;\mbox{for which}\;\;|e_0 \rangle = |e_x \rangle =
N^{-1} \mbox{$\sum\nolimits_x \! |x \rangle$}.  \label{sud5a}
\end{equation}
For $s$-fluctuations this model assumes the average time $\hat \tau_e$ to be
independent of $x$: $\hat \tau_e \equiv \tau_e$. Together with the model (%
\ref{sud5}) it implies equipopulated equilibration with $\hat q = 1$ so that
\begin{equation}
\hat {\mathcal{L}} = Q_{\!N}^{\!-1}(1\! -\! |e_x \rangle \langle e_x |) \; %
\mbox{and} \; \hat {\widetilde{W}}_{\!Q} = (1\!+\! Q_{\!N}\hat \Phi)^{\!-1}.
\label{sud6}
\end{equation}
It predicts, for example, for $n$-fluctuations ($W_i = W$)
\begin{equation}
\hat {\widetilde{\mathcal{R}}} = \hat {\widetilde{\mathcal{R}}}_n = \langle
\hat {\widetilde{P}}_{\!Q} \rangle [1 - \langle \hat {\widetilde{W}}_{\!Q}
\rangle]^{-1}.  \label{sud7}
\end{equation}

The model (\ref{sud5a}) clearly reveals all important specific features of
the non-Markovian-noise-induced relaxation in the simplest form.

\subsubsection{Diffusion model}

Another simple model allowing for analytical consideration of problem is the
diffusion model. In this work the diffusion model will be applied to the
analysis of recombination of the pair of radicals assuming that one of
radicals undergoes isotropic diffusion in three dimensional space, say $\{%
\mathbf{r}\}$-space, while another radical does not move and is located at $%
\mathbf{r} = 0$. Conventionally, the diffusive motion of the moving radical
is described by the Smoluchowski-type jump operator $\hat {\mathcal{L}} = 1
- \hat {\mathcal{P}}$ \cite{Shu1}. For simplicity we will consider isotropic
processes for which
\begin{equation}
\hat {\mathcal{L}} = \hat {\mathcal{L}}_D = - \lambda^2 r^{-2}
\nabla_{r}[r^{2}(\nabla_{r} + \nabla_{r}u_{r})],  \label{dif1}
\end{equation}
where $r = |\mathbf{r}|$, $\nabla_{r} = \partial/\partial r$ is the gradient
operator, $\lambda^2$ is the average square of the jump length independent
of ${r}$, and $u_{\mathbf{r}}$ is the external interaction potential.

The potential $u_r$ is assumed to be of the shape of (deep) spherically
symmetric square well with the radius $R$ much larger than the distance $d$
of closest approach: $u_r = -u_0 \theta (R - r)$ with $u_0 \gg 1$.

It worth noting that within the continuum model of stochastic jumps
resulting in $V(t)$- and $K (t)$-fluctuations the corresponding operators $%
\hat {\mathcal{V}}$ and $\hat {\mathcal{K}}_r$ are just functions of $%
\mathbf{r}$. In the considered spherically symmetric case
\begin{equation}
\hat {\mathcal{V}} = \sum\nolimits_r |r\rangle \hat V_r \langle r | \;\;%
\mbox{and} \;\; \hat {\mathcal{K}} = \sum\nolimits_r |r\rangle \hat K_r
\langle r |.  \label{dif2}
\end{equation}

\subsection{Short correlation time limit (SCTL)}

In practical applications of special importance is the SCTL for $V (t)$%
-fluctuations in which eq. (\ref{sud3}) can markedly be simplified. It
corresponds to the large characteristic (correlation) rate $w_c$ of
fluctuations, i.e. the large characteristic time of the dependence $\hat
\Phi (\hat \Omega)$: $w_c \gg \| V \|$. In our analysis we will discuss the
general class of models for which $\hat \Phi (\hat \Omega)$ is represented
in the form $\hat \Phi (\hat \Omega) \equiv \hat \Phi (\hat \Omega/\hat w)$.
For such models the correlation rate can naturally be defined as $w_c = \|
\hat w \|$.

In the SCTL the relaxation kinetics is described by the first terms of the
expansion of $\hat \Phi (\hat \Omega/\hat w)$ in small $\hat \Omega /w_c $,
since $\hat \Phi (\epsilon)$ is the increasing function of $\epsilon$ with $%
\hat \Phi (\epsilon) \overset{\epsilon \to 0}{\longrightarrow} 0$ [see Sec.
IIB and, in particular, eq. (\ref{ma17d})]. This fact allows one to
significantly simplify the problem under study. Nevertheless, some important
general conclusions can be made independently of the form of $\hat \Phi
(\Omega)$ as it will be shown in Sec. V.

\subsection{Models for quantum evolution.}

Here we describe two systems which will be analyzed to illustrate the
obtained general results: (1) the isolated quantum two-level system (TLS),
whose relaxation results from $V(t)$-fluctuations described within the
stochastic two-state model, and 2) the geminate pair of radicals diffusing
in the potential well in which spin evolution, affected by the fluctuating
reactivity $K(t)$, in turn strongly influences the reactivity.

\subsubsection{Isolated quantum TLS}

\paragraph{Two-level model.}

Quantum evolution of the TLS is governed by the hamiltonian (assumed to be
real matrix) with
\begin{equation}
H_s = \mbox{$\frac{1}{2}$}\omega_s \sigma_z \;\; \mbox{and} \;\; V = V_z
\sigma_z + V_x \sigma_x,  \label{two0}
\end{equation}
where
\begin{equation}
\sigma_z = \left[%
\begin{array}{rr}
1 & 0 \\
0 & -1 \\
&
\end{array}%
\right] \;\;\mbox{and} \;\; \sigma_x = \left[%
\begin{array}{cc}
0 & 1 \\
1 & 0 \\
&
\end{array}%
\right] \:
\begin{array}{c}
|+\rangle \\
|-\rangle \\
\end{array}%
.  \label{two1}
\end{equation}

\paragraph{Two-state model of fluctuations.}

In the two-state model of fluctuations $V (t)$-modulation is assumed to
result from jumps between two states (in $\{x \}$-space), say, $|x_+\rangle$
and $|x_-\rangle$. It is important to note that in the particular case of
two states any CTRWA-based kinetics model reduces to the simple two-state
SRM [see eqs. (\ref{sud5}) and (\ref{sud5a})] in which $\hat w \equiv
w,\:Q_2 = 1/2$, and
\begin{equation}
\hat {\mathcal{L}} = 2 (1\! - \!|e_x \rangle \langle e_x |) \;\:\mbox{with}
\;\: | e_x \rangle = \mbox{$\frac{1}{2}$}(|x_+\rangle \!+\! |x_-\rangle).
\label{two2}
\end{equation}

\paragraph{Simple variants of TLS and two-state model.}

Below we will consider two examples of these models:

1) \underline{Diagonal noise} \cite{And}: $\omega_s = 0$, $\, \mathcal{V}_x
= 0, \,$ and $\mathcal{V}_z = \omega_0 (|x_+\rangle \langle x_+| -
|x_-\rangle \langle x_-|)$, therefore
\begin{equation}
H_{\nu = \pm} = \pm \mbox{$\frac{1}{2}$}\omega_0 (|+\rangle \langle+|\, -
\,|-\rangle \langle-|);  \label{two3}
\end{equation}

2) \underline{Non-diagonal noise}: $\mathcal{V}_z = 0\:$ and $\: \mathcal{V}%
_x = v (|x_+\rangle \langle x_+| - |x_-\rangle \langle x_-|)$, so that
\begin{equation}
H_{\nu = \pm} = H_s \,\pm\, v (|+\rangle \langle-|\, + \,|-\rangle
\langle+|).  \label{two4}
\end{equation}

In our further analysis the first model is applied to the description of
dephasing while the second one is used in studying population relaxation.

\paragraph{Calculated observables.}

In the model (\ref{two1}) dephasing and population relaxation can be
characterized by two functions:

1) The spectrum $I (\omega)$ which is taken in the form corresponding to
Fourier transformed free-induction-decay (FTFID) experiments \cite{Wil}
\begin{equation}
I (\omega) = \mbox{$\frac{1}{\pi}$}\mathrm{Re}\langle s | \hat {\widetilde{%
\mathcal{R}}} (i\omega) |s\rangle;  \label{two5}
\end{equation}

2) The difference of level populations
\begin{equation}
N(t) = \langle n | \hat {\mathcal{R}} (t) |n\rangle.  \label{two5a}
\end{equation}

In these two functions
\begin{equation}
|s \rangle = \mbox{$\frac{1}{\sqrt{2}}$}\big||\!+\!- \rangle + |\!-\!+
\rangle \! \big\rangle \;\mbox{and}\; |n \rangle = \mbox{$\frac{1}{%
\sqrt{2}}$}\big||\!+\!+ \rangle - |\!-\!- \rangle \!\big\rangle.
\label{two5b}
\end{equation}

\subsubsection{Model for reactive radical pairs}

\paragraph{Magnetic field effects.}

The kinetics of RP recombination is known to be markedly affected by the RP
spin evolution which is controlled by the spin Hamiltonian $H$ of the pair.
The dependence of the recombination kinetics on RP spin state results in a
large number of phenomena called magnetic field effects \cite{St}.

In this work we will restrict ourselves to some simple and representative
effects observed in strong magnetic field $\mathbf{B}$ for which the Zeeman
interaction is much larger than the intraradical magnetic interactions
(hyperfine interaction, etc.). We will also consider the effect of the
external microwave field $B_1$ rotating with the frequency $\omega$ in the
plane perpendicular to the vector $\mathbf{B}$.

For strong magnetic fields and in the presence of the field $B_1$ the spin
Hamiltonian governing spin evolution of electrons in the pair of radicals,
say $a$ and $b$, can conveniently be written in the frame of reference
rotating together with $B_1$ with the frequency $\omega$ \cite{St}:
\begin{equation}
H_z = H_a + H_b \;\mbox{with} \; H_{\mu} = (\omega_{\mu}\! - \!\omega)
S_{\!\mu_z}\! + \omega_1 S_{\!\mu_x},  \label{rp1}
\end{equation}
where $\mu = a, b\,$ and $\omega_{\mu} = g_{\mu}\beta B + \sum_j A_j^{\mu}
I_{jz} $ is the Zeeman frequency of the radical $\mu$ possessing some
paramagnetic nuclei with hyperfine interactions $A_j^{\mu}$ and $\omega_1
\approx \frac{1}{2} (g_a + g_b)\beta B_1$. In case of need the matrix
representation of the Hamiltonian (\ref{rp1}) can be determined either in
the bases of radical spin states $|\pm\rangle_a |\pm\rangle_b$ or in the
basis of eigenstates of the total electron spin $\mathbf{S} = \mathbf{S}_a +
\mathbf{S}_b$: singlet ($|S\rangle$) and triplet ($|T_{0, \pm}\rangle$)
states, which are expressed as $|S\rangle = \mbox{$\frac{1}{\sqrt{2}}$}%
(|+\rangle_a |-\rangle_b - |-\rangle_a|+\rangle_b), \, |T_0\rangle = %
\mbox{$\frac{1}{\sqrt{2}}$}(|+\rangle_a |-\rangle_b +
|-\rangle_a|+\rangle_b), \,$ and $\, |T_{\pm}\rangle = |\pm \rangle_a |\pm
\rangle_b$;

The RP recombination can be treated as a contact reaction at a distance of
closest approach $d$ using the simple model \cite{St}
\begin{equation}
\hat K_r = k_0 \hat \kappa_S \theta (r - d)\theta (d+\Delta - r),
\label{rp2}
\end{equation}
where
\begin{equation}
\hat \kappa_S = \{\mathcal{P}_s, \dots \}  \label{rp2a}
\end{equation}
is anticommutator ($\{\mathcal{P}_s, \rho \}= \mathcal{P}_s \rho + \rho
\mathcal{P}_s$) in which $\mathcal{P}_s = |S \rangle \langle S |$ is the
operator of projection on the singlet ($|S\rangle$) spin state of RP.

The RP is assumed to be initially created in the singlet state $|S \rangle$
within the potential well at a distance $r = r_i < R\:$ ($r_i > d$), so that
\begin{equation}
\rho (r,t=0) \equiv \rho_i (r) = (4\pi r_i^2)^{-1}\delta (r-r_i) \mathcal{P}%
_s.  \label{rp3}
\end{equation}

\paragraph{Observables.}

In experiments on magnetic field effects a number of observables are
discussed \cite{St}. Here we analyze the most simple ones:

1) magnetically affected reaction yield (MARY) \cite{St}, measured in the
external constant magnetic field,

2) reaction yield detected magnetic resonance (RYDMR) \cite{St}, i.e.
microwave field influenced recombination yield.

In both types of experiments the observables under study are recombination ($%
Y_r$) and dissociation ($Y_d$) yields:
\begin{equation}
Y_r = (d/r_i)^2\Delta k_0 \mathrm{Tr}[\mathcal{P}_s \hat {\widetilde {%
\mathcal{G}}} (d,r_i|\epsilon = 0)\mathcal{P}_s]  \label{rp4}
\end{equation}
and $Y_d = 1 - Y_r$.

Naturally, the expression (\ref{rp4}) should be averaged over nuclear
configurations (over $\omega_{\mu}$). However, in our further discussion we
will omit this evident procedure and analyze the behavior of $Y_r$ for fixed
$\omega_a$ and $\omega_b$ (note that the case of fixed $\omega_a$ and $%
\omega_b$ can be realized experimentally, for example, with RPs which do not
contain paramagnetic nuclei).

\section{Results and discussion}

\subsection{Isolated quantum TLS}

\subsubsection{Some general results in the SCTL}

\paragraph{Small $\|H_s\|/ w_c \ll 1$.}

Within the SCTL relatively simple and general results can be obtained in the
case $\|H_s\|/ w_c \ll 1$. In the lowest order in $\|\hat \Phi (\hat
\Omega/w_c)\| \ll 1$ \cite{Shu4}
\begin{equation}
\hat {\widetilde{\mathcal{R}}} \approx \hat {\widetilde{\mathcal{R}}}_n
\approx \langle \hat w^{\alpha}{\hat \Omega}^{\!-1} \hat \Phi (\hat
\Omega)\rangle / \langle \hat w^{\alpha} \hat \Phi (\hat \Omega) \rangle.
\label{sctl1}
\end{equation}
This formula holds for any initial matrix $\hat {\widetilde{W}}_i$ and, in
particular, for $s$-fluctuations, if $\|\hat \tau_e \| \sim 1/w_c \ll
1/\|\hat \Omega \|$. It can easily be obtained from eq. (\ref{sud3}) if one
takes into account that $\hat q^{-1}\hat Q = N_0 \hat w^{\alpha}$.

\paragraph{Large $\|H_s\|/ w_c \gtrsim 1$.}

Somewhat more complicated SCTL-case $\| H_s \|/w_c \simeq 1$ can be analyzed
by expanding $\hat {\widetilde {\mathcal{G}}}$ in powers of the parameter $%
\xi = \| V \|/\| H_s \| \ll 1$. In particular, within the general two-level
model [eq. (\ref{two1})] with $V_d = 0$ in the second order in $\xi$ the
diagonal and non-diagonal elements of $\rho (t)$ are decoupled and the
corresponding elements of $\hat {\mathcal{R}} (t)$ are expressed in terms of
the universal function
\begin{eqnarray}
\mathcal{P}_{k} (t) &=& \frac{1}{2\pi i}\int_{-i\infty}^{i\infty}\!d\epsilon
\,\frac{e^{i\epsilon t}}{\epsilon + k \epsilon/\langle \hat \Phi
(\epsilon)\rangle}:  \label{sctl4a} \\
\langle \mu |\hat {\mathcal{R}}(t)|\mu \rangle &=& e^{-i\omega_{\mu} t}
\mathcal{P}_{k_{\mu}} (t), \;\;(\mu =n, +-,\, -+ ),  \label{sctl4b}
\end{eqnarray}
where
\begin{equation}
\omega_{\mu} = \langle \mu |\hat { H}_{s}|\mu \rangle,\;\; k_n = 2\mathrm{Re}
(k_{+-}),  \label{sctl15}
\end{equation}
and
\begin{equation}
k_{+-} = k_{-+}^{*} = \mbox{$\frac{1}{2}$}\omega_s^{-2} \langle \mathcal{V}%
_n\hat q^{-1}[1-\hat {\widetilde W}_Q (2i\omega_s)] \mathcal{V}_n \rangle.
\label{sctl2}
\end{equation}

\subsubsection{Anomalous $V(t)$-fluctuations}

The simplest model for anomalous fluctuations can be written as \cite{Met}
\begin{equation}
\hat \Phi (\epsilon) = (\epsilon/\hat w)^{\alpha}, \; (0 < \alpha < 1),
\label{res1}
\end{equation}
where $\hat w$ is the matrix of characteristic fluctuation (correlation)
rates diagonal in $|x\rangle$-basis. For the sake of simplicity, $\hat w$ is
assumed to be independent of $x$, i.e. $\hat w \equiv w (= w_c)$ so that one
can use formula (\ref{sud7}) for evaluation of $\hat {\widetilde{\mathcal{R}}%
}_n$. The model (\ref{res1}) describes anomalously slow decay of the
PDF-matrix $\hat W (t) \sim 1/t^{1+\alpha}$ (very long memory effects in the
system \cite{Met}), for which only the case of non-stationary ($n$)
fluctuations is physically sensible.

For small $\|H_s\|/ w_c \ll 1$ [see eq. (\ref{sctl1})] the model (\ref{res1}%
) yields the expression constituting the important result of the work:
\begin{equation}
\hat {\widetilde{\mathcal{R}}}_n = \langle \hat \Omega^{\alpha -
1}(\epsilon) \rangle \langle \hat \Omega^{\alpha}(\epsilon) \rangle^{-1} \; %
\mbox{with} \;\, \hat \Omega (\epsilon) = \epsilon + i\hat {\mathcal{H}}.
\label{res2}
\end{equation}
This expression demonstrates the surprising property of relaxation induced
by anomalous noise in the SCTL: the evolution operator $\hat {\widetilde{%
\mathcal{R}}}_n (\epsilon)$ [and $\hat {\mathcal{R}}_n (t)$], which
determines relaxation kinetics, is independent of the characteristic rate $w$%
.

It is also worth noting that for $\alpha = 0$ and $\alpha = 1$ formula (\ref%
{res2}) describes relaxation kinetics corresponding to the static and
fluctuation narrowing limits \cite{Abr} in which
\begin{equation}
\hat {\widetilde{\mathcal{R}}}_n = \langle \hat \Omega^{- 1} (\epsilon)
\rangle \;\;\mbox{and} \;\; \hat {\widetilde{\mathcal{R}}}_n = 1/\langle
\hat \Omega (\epsilon) \rangle,  \label{res2_1}
\end{equation}
respectively. At intermediate values $0 < \alpha < 1$ the kinetics is
represented by a non-trivial combination of the static- and narrowing-like
expressions whose relative contribution is determined by $\alpha$, as it is
seen from eq. (\ref{res2}).

Of certain interest is the limit $\alpha \to 1$ in which formula (\ref{res2}%
) predicts the Bloch-type exponential relaxation:
\begin{equation}
\hat {\widetilde{\mathcal{R}}}_n (\epsilon) \approx [\epsilon + i \hat {H}_s
+ (\alpha -1)\langle \hat {\Omega} \ln (\hat {\Omega})\rangle ]^{-1}.
\label{res2a}
\end{equation}
The relaxation is controlled by the rate matrix $\hat W_r = (\alpha-1)%
\mathrm{Re}\langle \hat {\Omega} \ln (\hat {\Omega})\rangle$, and is
accompanied by frequency shifts represented by $\hat h = i(\alpha -1)\mathrm{%
Im}\langle \hat {\Omega} \ln (\hat {\Omega})\rangle$. The peculiarity of
anomalous relaxation, however, shows itself in the independence of matrices $%
\hat W_r$ and $\hat h$ (unlike those in the conventional Bloch equation)
from the characteristic rate $w$ of $V(t)$-fluctuations.

\subsubsection{Anomalous dephasing for diagonal noise}

In the simple model (\ref{two3}) of diagonal noise the spectrum $I(\omega)$
can be obtained in analytical form in relatively general assumptions on $%
V(t) $-fluctuations.

\paragraph{General SRM and SCTL.}

In the SCTL (i.e. for large rate $w$) within the general SRM (\ref{sud1})
relatively simple expression for $I(\omega)$ can be derived without any
assumption on the structure of energy levels (for any number of states) \cite%
{Shu4}:
\begin{equation}
I(\omega) = \frac{\sin\varphi}{\pi} \frac{ \psi_{-}^{\alpha}
\psi_{+}^{\alpha-1} + \psi_{-}^{\alpha-1} \psi_{+}^{\alpha}}{%
(\psi_{-}^{\alpha})^2 + (\psi_{+}^{\alpha})^2 + 2 \psi_{-}^{\alpha}
\psi_{+}^{\alpha} \cos \varphi},  \label{res3}
\end{equation}
where
\begin{equation}
\varphi = \pi \alpha \;\;\mbox{and}\;\; \psi_{\pm}^{\beta} = \langle |\omega
- 2V_d |^{\beta} \theta [\pm (\omega - 2V_d)]\rangle  \label{res3a}
\end{equation}
with $\theta (z)$ being the Heaviside step-function. In general, this
formula is too cumbersome for studying the specific features of the
spectrum. Much more clearly they can be revealed with the use of the
two-state SRM (see below).

\paragraph{Arbitrary rate $w$ in two-state SRM.}

The two-state SRM (\ref{two2}) allows for the analytical analysis of the
spectrum $I(\omega)$ for any value of $w$, i.e. outside the region of
applicability of the SCTL:
\begin{equation}
I(\omega) = \frac{2\xi}{\pi\omega_0}\frac{\sin\phi_0}{z_{+}z_{-}(\eta +
\eta^{-1} + 2\cos\phi_0)}.  \label{res4c}
\end{equation}
In this expression
\begin{eqnarray}
z_{\pm} &=& \xi (1 \pm \omega/\omega_0), \;\;\mbox{with}\;\; \xi =
\omega_0/(2^{1/\alpha}w), \quad\quad\quad \\
\eta \; &=& \zeta (z_{+})/\zeta(z_{-}) \;\;\mbox{and} \;\; \phi_0 =
\phi(z_{+}) + \phi(z_{-}),\;\;  \label{res4d}
\end{eqnarray}
where
\begin{eqnarray}
\zeta (z) &=& |z|^{\alpha}/\sqrt{1 + |z|^{2\alpha} +
2|z|^{\alpha}\cos(\varphi/2)}, \\
\phi (z) &=& \mathrm{sign} (z)\arctan \!\bigg[\frac{\sin(\varphi/2)}{%
|z|^{\alpha}+\cos (\varphi/2)}\bigg].\;\;\;  \label{res4e}
\end{eqnarray}

Figures 1 and 2 demonstrate a large variety of dependencies $I(\omega)$ for
different values of the parameter $\xi = \omega_0/(2^{1/\alpha}w)$. In
general, the spectrum possesses two peaks at $\omega = \pm \omega_0$ which
at $\xi \ll 1$ and $\alpha \to 1$ collapse into one central peak (see next
paragraph). Since in accordance with the definition (Sec. IVB) the limit $%
\xi \ll 1$ corresponds to the SCTL, the collapse of lines for $\alpha \to 1$
looks quite natural \cite{Abr}. Simple analysis of eq. (\ref{res4c}) and
calculation results displayed in Fig. 1 show that in this limit general
spectrum (\ref{res4c}) reduces to the limiting one confined in the region $%
|\omega|/\omega_0 < 1$ [see eq. (\ref{res4}) and its discussion]. With the
increase of $\xi$, however, the delocalization of the spectrum outside this
region is predicted (Fig. 2). The delocalization becomes especially
pronounced in the limit $\alpha \to 1$, as expected.

\paragraph{SCTL within two-state SRM.}

In the two-state SRM (\ref{two2}) formula (\ref{res3}) is significantly
simplified to give the same result as eq. (\ref{res4c}) in the limit $\xi
\ll 1$:
\begin{equation}
I(\omega )=\frac{\sin \varphi }{2\pi \omega _{0}}\theta (y)\frac{y+y^{-1}+2}{%
y^{\alpha }+y^{-\alpha }+2\cos \varphi }  \label{res4}
\end{equation}%
where $y=(\omega _{0}+\omega )/(\omega _{0}-\omega )$ (see also ref. \cite%
{Silb}). According to this formula anomalous dephasing (unlike conventional
one \cite{Abr}) leads to broadening of $I(\omega )$ only in the region $%
|\omega |<\omega _{0}$ and singular behavior of $I(\omega )$ at $\omega
\rightarrow \pm \omega _{0}$: $I(\omega )\sim 1/(\omega \pm \omega
_{0})^{1-\alpha }$. For $\alpha >\alpha _{c}\approx 0.59$ [$\alpha _{c}$
satisfies the relation $\alpha _{c}=\cos (\pi \alpha _{c}/2)$] the formula
also predicts the occurrence of the central peak (at $\omega =0$) \cite{Silb}
of Lorenzian shape and width
\begin{equation}
w_{L}\approx \omega _{0}\cos (\varphi /2)/\sqrt{\alpha ^{2}-\cos
^{2}(\varphi /2)}:  \label{res4a}
\end{equation}%
\begin{equation}
I(\omega )\approx (2\pi \omega _{0})^{-1}\tan (\varphi /2)/[1+(\omega
/w_{L})^{2}],  \label{rs4b}
\end{equation}%
whose intensity increases with the increase of $\alpha -\alpha _{c}$ (see
Fig. 1a). At $\alpha \sim 1$ the parameters of this peak are reproduced by
eq. (\ref{res2a}) in which $\langle \hat{\Omega}\ln (\hat{\Omega})\rangle
=-(\pi /2)\omega _{0}$. The origination of the peak indicates the transition
from the static broadening at $\alpha \ll 1$ to the narrowing one at $\alpha
\sim 1$ [see eq. (\ref{res2})].

It is worth noting that, of course, for systems with complex spectra this
transition can strongly be smoothed and almost indistinguishable
experimentally.

\subsubsection{Anomalous dephasing for non-diagonal noise}

\paragraph{Dephasing in SCTL for $\|H\|/w \ll 1$.}

The two-state SRM (\ref{two2}) enables one to analyze the behavior of the
spectrum $I (\omega)$ in the complicated case $\xi_s = \omega_s/w, v/w \ll 1$
(within the applicability region for the SCTL). In this case dephasing is
determined by eq. (\ref{res2}). After some manipulations one arrives at
\begin{equation}
I(\omega) = \frac{1}{2\pi} \mathrm{Re} \bigg[ \frac{\Omega_{-1}(i\omega) +
r_0^2(i\omega)^{\alpha-1} \Omega_{-\alpha}(i\omega)}{1 + r_0^2
(i\omega)^{\alpha}\Omega_{-\alpha} (i\omega)}\bigg],  \label{deph1}
\end{equation}
where
\begin{eqnarray}
&&\Omega_{\beta}(\epsilon) = \mbox{$\frac{1}{2}$}[(\epsilon\! +\! 2i\!
E_0)^{\beta}+(\epsilon \!- \!2i\! E_0)^{\beta}],  \label{deph2} \\
&&r_0 = 2v/\omega_s, \;\; \mbox{and} \;\; E_0 = (\omega_s/2) \sqrt{1+r_0^2}.
\label{deph2a}
\end{eqnarray}

The spectrum $I (\omega) $ predicted by eq. (\ref{deph1}) is depicted in
Fig. 3 as a function of $z = \omega/E_0 $ for two values of $\alpha$ and
different values of the parameter $r$. This figure demonstrates non-trivial
specific features of the shape of $I (\omega)$ depending on the values of $%
\alpha$ and $r$. First of all, similarly to the case of diagonal noise, (in
the limit $\xi_s = \omega_s/w, v/w \ll 1$) the spectrum $I (\omega)$ is
localized in the region $|\omega | < E_0$ for all values of $\alpha$.
Outside this region $I (\omega) = 0$.

The behavior of $I (\omega)$ significantly changes with the increase of $%
\alpha$:

1) At small $\alpha \lesssim 0.6$ the spectrum consists of three peaks (see
Fig. 3a): the central peak at $\omega = \omega_c = 0$ and symmetric edge
peaks at $\omega = \omega_{\pm} = \pm E_0$ (at the spectrum edges), in
vicinity of which the behavior of $I (\omega) \sim 1/|\omega -
\omega_{\mu}|^{1-\alpha}$, where $\mu = c, \pm $. The intensity of the
central peak decreases as $\alpha$ is increased.

2) For large $\alpha \gtrsim 0.6$ each of two symmetric peaks at $\omega =
\pm E_0$ split into two ones, so that the spectrum possesses five peaks: at $%
\omega = \omega_c = 0$, at $\omega = \pm E_0$ and at $\omega = \pm \omega_a$
with $\omega_a < E_0$ (see Fig. 3b). Moreover, the intensity of three
original edge and central peaks (at $\omega = \pm E_0$ and $\omega = 0$)
decreases as $\alpha \to 1$ while the intensity two additional peaks
increases. Besides, the two peaks at $\omega = \pm \omega_a$ approach each
other, i.e. $\omega_a \to 0$ with the increase of $r$ (as it is shown in
Fig. 3b), and in the limit $r \gg 1$ collapse into one peak.

\paragraph{Dephasing in SCTL for large $\protect\omega_s/w \gtrsim 1$.}

The model (\ref{two4}) reveals some additional specific features of the
kinetics of phase relaxation in the case of not very large $w$, when $\xi_s
= \omega_s/w \gtrsim 1$. For example, as it is seen from eqs. (\ref{sctl4a})
and (\ref{sctl4b}), in the limit $\|H_s\| \sim \omega_s \gtrsim w$ matrix
elements $\langle \mu |\mathcal{R} (t) | \mu \rangle ,\,(\mu = +-,\,-+),$
which describe phase relaxation are given by
\begin{equation}
\langle \mu |\mathcal{R} (t) | \mu \rangle = e^{-i\omega_{\mu} t }
E_{\alpha} [-k_{\mu}(w t)^{\alpha}],  \label{res50}
\end{equation}
where
\begin{equation}
E_{\alpha} (-z) = \frac{1}{2\pi i}\int_{-i\infty}^{i\infty}\! dy\, \frac{%
e^{y}}{y + z y^{1-\alpha}}  \label{res51}
\end{equation}
is the Mittag-Leffler function \cite{Bat}. In this case the spectrum
\begin{equation}
I (\omega) = I_0 (\omega_{+}) + I_0 (\omega_{-}), \;\;\mbox{where}
\;\;\omega_{\pm} = \omega_s \pm \omega  \label{res5}
\end{equation}
and
\begin{equation}
I_0 (\omega) = \frac{\sin{\varphi}}{\pi} \frac{\sin \phi_n}{|x|(|x|^{\alpha}
+ |x|^{-\alpha} + 2\cos \phi_n)}  \label{res5a}
\end{equation}
with $\varphi = \pi \alpha$,
\begin{equation}
x = \omega/(|k_{+ -}|^{1/\alpha}w), \;\; n_0 = (\pi |k_{+
-}|^{1/\alpha}w)^{-1},  \label{res5b}
\end{equation}
and
\begin{equation}
\phi_n = \frac{\varphi}{2} + \mathrm{sign} (x)\arctan\!\! \Big[\frac{%
2^{\alpha^{-1}}\sin (\mbox{$\frac{1}{2}$}\varphi)}{2^{\alpha^{-1}}\cos(%
\mbox{$\frac{1}{2}$}\varphi)+\omega_s/w}\Big].  \label{res5c}
\end{equation}

Formulas (\ref{res5})-(\ref{res5c}) predict singular behavior of $I (\omega)$
at $\omega \sim \pm\omega_s$: $I(\omega) \sim
1/|\omega\pm\omega_s|^{1-\alpha}$, and slow decrease of $I (\omega)$ with
the increase of $|\omega \pm \omega_s|/\kappa \gg 1$: $I(\omega) \sim
1/|\omega \pm \omega_s|^{1+\alpha}$.

In the limit $\xi_s = \omega_s/w \ll 1\:$ $\phi (x) \approx \pi\alpha \theta
(x)$ so that $I_0 (\omega) \sim \theta (\omega)$. This means that for $\xi_s
\ll 1\:$ the spectrum $I (\omega)$ is localized in the region $|\omega | <
\omega_s$ and looks similar to $I (\omega)$ for diagonal dephasing at $%
\alpha < \alpha_c$ and $\xi \ll 1$ (see Fig. 1a). For $\xi_s \gtrsim 1$,
however, $I (\omega)$ is non-zero outside this region as well, moreover, in
the limit $\xi_s \gg 1$ the spectrum $I_0 (\omega)$ becomes symmetric: $I_0
(\omega) = I_0 (-\omega)$. Such dependence of $I (\omega)$ on $\xi_s$ is
also very similar to $I (\omega)$-dependence on $\xi$ found above in the
case of diagonal noise.

It is interesting to note that for $\xi_s \ll 1\,$ functions $\langle \mu |%
\mathcal{R} (t) | \mu \rangle$ and $I(\omega)$ are independent of $w$ [in
agreement with eq. (\ref{res2})] since $k_{\mu} \sim (\omega_s/w)^{\alpha}$
and $k_{\mu}(w t)^{\alpha} \sim (\omega_s t)^{\alpha}$. In the opposite
limit, however, $k_{\mu} \sim w^{0}$ so that the characteristic relaxation
time $\sim w^{-1}$.

\subsubsection{Anomalous population relaxation}

Specific features of anomalous population relaxation can be analyzed with
the model of non-diagonal noise (\ref{two4}).

In particular, in the limits $\|H_s\| \sim \omega_s \gtrsim w$ and $1-\alpha
\ll 1$ with the use of eqs. (\ref{sctl4a}),(\ref{sctl4b}) and (\ref{res2a})
one gets
\begin{equation}
N(t) = E_{\alpha} [-k_n(w t)^{\alpha}] \;\;\mbox{and} \;\; N(t) = e^{-
w_{\alpha} t},  \label{res6}
\end{equation}
respectively, where $E_{\alpha} (-x)$ is the Mittag-Leffer function defined
above and $w_{\alpha} = k_n (\alpha\to 1)w \sim 1-\alpha$. The first of
these formulas predicts very slow population relaxation at $t > \tau_r =
w^{-1}(k_n/w)^{1/\alpha}$: $N (t) \sim 1/t^{\alpha}$. Similar to $I(\omega)$
the function $N (t)$ is, in fact, independent of $w$ for $\xi_s = \omega_s/w
\ll 1$ because in this limit $k_n \sim (\omega_s/w)^{\alpha}$. In the
opposite limit $\xi_s > 1$ the characteristic time population relaxation is $%
\sim w^{-1}$ since $k_n$ is independent of $w$ as in the case of phase
relaxation.

For $\|H_s\|, \|V\| \ll w$ one obtains \cite{Shu4}
\begin{equation}
N (t) = \frac{1}{2\pi i} \int_{-i\infty}^{i\infty}\!\! d\epsilon\,
e^{\epsilon t}\,\frac{ \epsilon^{\alpha-1} + r_0^2\Omega_{\alpha-1}(\epsilon)%
}{\epsilon^{\alpha} + r_0^2\Omega_{\alpha}(\epsilon)},  \label{res7}
\end{equation}
where $\Omega_{\beta}(\epsilon)$ and $r_0$ are defined in eqs. (\ref{deph2})
and (\ref{deph2a}), respectively.

It is easily seen that in the corresponding limits the expression (\ref{res7}%
) reproduces formulas (\ref{res6}) with $k_n \approx
2^{\alpha-1}\cos(\pi\alpha/2)(E_0/w)^{\alpha}$ and $w_{\alpha} \approx \pi
(1-\alpha)v^2/E_0$. Outside these limits $N (t)$ can be evaluated
numerically (some results are shown in Fig. 4). In general, $N (t)$ is the
oscillating function (of frequency $\sim E_0$) with slowly decreasing
average value and oscillation amplitude: for $\tau = E_0 t \gg 1$ $N (\tau)
\sim 1/\tau^{\alpha}$ (except the limit $\alpha \to 1$). At large $\tau =
E_0 t$ one can estimate the asymptotic behavior of $N (t)$:
\begin{eqnarray}
N (\tau) &\approx& \frac{2}{\pi}\Gamma (\alpha)\Big[2^{\alpha}\sin (%
\mbox{$\frac{1}{2}$}\varphi) r_0^{-2}  \notag \\
&&+\sin\varphi \frac{2^{-\alpha}r_0^2}{2+2^{\alpha}r_0^2} \cos (2\tau + %
\mbox{$\frac{1}{2}$}\varphi)\Big] \frac{1}{\tau^{\alpha}}.\quad
\label{res7a}
\end{eqnarray}

\subsubsection{Fractional Bloch equation}

The kinetic dependencies found in this section for $\|V\|/\|H_s\| \ll 1$ and
expressed in terms of the Mittag-Leffler function are conveniently
represented in the form of the equation similar to the conventional Bloch
equation for the density matrix but with the fractional derivatives. It is
easily seen that the kinetic functions (\ref{res50}) and (\ref{res6}) can be
considered as solution of the equation
\begin{equation}
d{\hat {\mathcal{R}}}/dt =-i\hat H_s \hat {\mathcal{R}} - w^{\alpha} {\hat k}
[{}_{0}\! \hat D_{\!t}^{\!1\!- \alpha}]_{s}^{\!\times} \hat {\mathcal{R}},
\label{res8}
\end{equation}
where
\begin{equation}
[{}_{0}\! \hat D_{\!t}^{\!1\!- \alpha}]_{s}^{\!\times}\!f = \frac{1}{\Gamma
(\alpha)} e^{-i\hat H_s t}\frac{\partial}{\partial t} \! \int_0^t \!\! d\tau
\frac{e^{i\hat H_s \tau}}{(t\!-\!\tau)^{1\!-\alpha}} f (\tau)  \label{res9}
\end{equation}
is the modified Liouville-Riemann fractional integral operator, and
\begin{equation}
\hat k = \mbox{$\sum\nolimits_{\mu}$} |\mu\rangle k_{\mu} \langle \mu | ,\;
\;\;(\mu = + -, - +, n ).  \label{res9b}
\end{equation}
This equation generalizes the well known classical expressions \cite{Met} to
the quantum processes.

\subsubsection{Weakly anomalous fluctuations}

The analysis presented above shows that the effect of the anomalous noise,
i.e. fluctuating interaction [whose correlation functions is anomalously
long tailed: $P (t) \sim t^{-\alpha}$ with $0 < \alpha < 1$], on quantum
systems can be very strong, manifesting itself in anomalous relaxation
kinetics. With the increase of $\alpha$ up to $\alpha > 1$ the effects of
anomaly of interaction fluctuations become weaker but, nevertheless, they
still manifest themselves in the relaxation kinetics.

To clarify these effects we will briefly discuss the model in which
\begin{equation}
\Phi (\epsilon) = (\epsilon /w) + \zeta (\epsilon /w)^{1+\alpha},
\label{res10a}
\end{equation}
where $0 < \alpha < 1$, and $w$ and $\zeta$ are the constants with $\zeta
\ll 1$ [small value of $\zeta$ ensures that $W (t) > 0$]. This $\Phi
(\epsilon)$ corresponds to the waiting time PDF-matrix $\hat W (t)$ for
which the average time $\hat \tau_e = \langle t \rangle = \int_0^{\infty} \!
dt \, t \hat W (t) = w^{-1}$ is finite but the higher moments $\langle t^n
\rangle$ with $n \geq 2$ do not exist.

Possible effects of this weakly anomalous noise can be analyzed within the
SCTL with the use of eqs. (\ref{sctl1})-(\ref{sctl4b}). For example, in the
limit $\|H\|/w \ll 1$ one obtains formula
\begin{equation}
\widetilde{\mathcal{R}} \approx [\epsilon + i \hat H_s + \zeta w^{-\alpha}
\langle (i\hat {\mathcal{H}})^{1+\alpha} - (i\hat
H_s)^{1+\alpha}\rangle]^{-1}.  \label{res10}
\end{equation}
which predicts the Bloch-type relaxation of both phase and population, but
with the rate
\begin{equation}
\hat W_r = \zeta w^{-\alpha} \mathrm{Re} \langle (i\hat {\mathcal{H}}%
)^{1+\alpha} - (i\hat H_s)^{1+\alpha}\rangle  \label{res11}
\end{equation}
depending on the $w$ as $w^{-\alpha}$, i.e. slower than in the conventional
Bloch equation ($\hat W_r \sim 1/w$ \cite{Abr}).

More detailed analysis also demonstrates that in this expression for $%
\widetilde{\mathcal{R}}$ the terms $\sim w (\epsilon/w)^{1+\alpha}$ are also
expected. They lead to the inverse power-type asymptotic behavior of $%
\langle \mu | \hat R (t)|\mu \rangle \sim 1/t^{2+\alpha}$ which, however, is
observed only at very long times $t \gg w^{-1}$. This conclusion can easily
be clarified with the use of the general formula (\ref{sctl4a}) valid in the
case $\|H_s \|/w \lesssim 1, \|V\|/\|H_s\| \ll 1$ for any model of $\Phi
(\epsilon)$. According to this formula for the model (\ref{res10a}) the
kinetics of both phase and population relaxation is similar in its
mathematical form to the probability of fluctuations $P (t)$ [see eq. (\ref%
{nmark1})] which is, evidently, of power-type behavior at long times: $P (t)
\sim 1/t^{2+\alpha}$, in agreement with the above statement.

Formula (\ref{sctl4a}) allows for making the following general conclusion on
the relaxation kinetics: the kinetics is of anomalous long tailed
inverse-power-type behavior for any CTRWA-based model of fluctuations
assuming singular $\Phi (\epsilon)$ with brunching points.

\subsection{Anomalous magnetic field effects on RP recombination}

In accordance with eq. (\ref{rp4}), the observables investigated in MARY and
RYDMR experiments are expressed in terms of the operator ${\widetilde {%
\mathcal{G}}} (r,r_i|\epsilon)$. Within the CTRWA equation for this operator
is determined by the model of migration of the mobile radical.

\subsubsection{Anomalous diffusion and anomalous SLE.}

Here we will consider the anomalous diffusion model of migration in which
the memory is anomalously long and is described by the operator $\hat \Phi
(\epsilon)$ (\ref{res1}). Notice that in the considered process of spatial
diffusion the (diagonal) matrix of characteristic rates $\hat w$ in the
operator $\hat \Phi (\epsilon)$ is actually represented as a distance
dependent function: $\hat w = \sum_r |r\rangle w_r \langle r |$. For
simplicity, in our further discussion of radical diffusion we assume the
rate $w_r$ and the exponent $\alpha_r$ to be independent of $r$: $w_r \equiv
w$ and $\alpha_r \equiv \alpha$. In this case the CTRWA, with $\hat \Phi
(\epsilon)$ given by (\ref{res1}), is known to predict anomalous diffusion
\cite{Met}.

The considered anomalous diffusion of the mobile radical of the RP,
evidently, results in non-stationary fluctuations of reactivity $\hat K_r$.
The effect of these fluctuation is described by the operator ${\widetilde {%
\mathcal{G}}} (r,r_i|\epsilon)$ satisfying the non-Markovian SLE (\ref{ma18}%
) which can be represented in terms of the so called fractional diffusion
equation (for the Laplace transform) as follows
\begin{equation}
\hat \Omega_r \hat {\widetilde {\mathcal{G}}} = - \hat {\mathcal{L}}_D (%
\hat {\widetilde M}_r {\widetilde {\mathcal{G}}}) + \delta (r-r_i),
\label{mfe1}
\end{equation}
where $\hat \Omega_r = \epsilon + \hat K_r + i\hat H$ and
\begin{equation}
\hat {\widetilde M}_r = \hat \Omega_r \Phi (\hat \Omega_r) = w(\hat
\Omega_r/w)^{1-\alpha}.  \label{mfe2}
\end{equation}

In general, solution of the non-Markovian SLE (\ref{mfe1}) with the
Smoluchowski-type $\hat {\mathcal{L}}_D$ [see eq. (\ref{dif1})] is a fairly
complicated (though, in principle, analytically tractable \cite{Shu5})
problem. The most interesting specific features of magnetic field effects,
however, can quite clearly be illustrated in some limiting cases allowing
for considerable simplification of the obtained general expressions. In our
analysis we will concentrate on one of such cases, corresponding to the
limit of the deep well $u_r$ and fast diffusion within the this well, which
is described by the cage model \cite{Shu5}.

\subsubsection{The anomalous-cage model.}

Similar to the case of conventional diffusion, fast anomalous diffusion in a
large enough and deep potential well $u_r = -u_0 \theta (R-r)$, for which $R
\gg d$ and $u_0 \gg 1$ leads to rapid relaxation of the initial
non-equilibrium spatial distribution of the radical within the well during
the time $\tau_D = (R/\lambda)^{2/\alpha}/w$ and formation of the nearly
homogeneous quasiequilibrium state (cage) within the well. At longer times $%
t > \tau_D$ reaction and dissociation are shown to result in the
quasistationary decay of this state which is, naturally, independent of the
distance $r_i$ of RP creation. At these times, also as for reactions
assisted by conventional diffusion, the (anomalous) kinetics of the process
under study is described by lowest pole in the expression (\ref{ma18})
(determined by the lowest eigenvalue of the operator $\hat {\mathcal{L}}_D$)
\cite{Shu5}. In what follows this approximation will be called the
anomalous-cage model.

Keeping this lowest pole after some algebraic manipulations similar to those
presented in ref. [27] one arrives at formula
\begin{equation}
Y_r = \mathrm{Tr}\bigg[\mathcal{P}_s {\hat l_r}\frac{1}{\hat l_r + l_d +
(i\hat H_z/w)^{\alpha}}\mathcal{P}_s\bigg],  \label{mfe3}
\end{equation}
where \cite{Shu2}
\begin{equation}
\hat l_r = \frac{\lambda^2}{R^2} \frac{(\Delta d/\lambda^2)(k_0\hat
\kappa_S/w)^{\alpha}}{1+(\Delta d/\lambda^2)(k_0\hat \kappa_S/w)^{\alpha}},
\;\; l_d = \frac{\lambda^2}{R^2} e^{-u_0},  \label{mfe4}
\end{equation}
and $\mathcal{P}_s = |S \rangle \langle S|$ is the operator of projection on
the singlet ($|S \rangle$) state of the RP.

Formula (\ref{mfe3}) is quite suitable for the analysis of the problem under
study. Unfortunately, in general, in the cage model (\ref{mfe3}) the
expressions for $Y_r$ are still fairly cumbersome and can be used mainly for
numerical estimations.

\subsubsection{Specific features of MARY and RYDMR}

To reveal characteristic properties of $Y_r$-dependence on the parameters of
the model we consider simple limiting cases which will be somewhat different
for MARY and RYDMR.

\paragraph{Analysis of MARY}

In the considered limit of strong magnetic field $B$, in which $g_{\mu}
\beta B \gg A_j^{\mu}$, the effect of spin evolution on reaction yield
called MARY can be studied within the $ST_0$-approximation. In this
approximation, which takes into account that in the strong magnetic field
limit the contribution of $|T_{\pm}\rangle $-terms to the reaction yield is
negligibly small, the RP spin Hamiltonian is written as
\begin{equation}
H_z^{0} = \mbox{$\frac{1}{2}$} \delta\omega (|S\rangle \langle T_0|\! +
\!|T_0\rangle \langle S| ) \;\mbox{with} \; \delta\omega = \omega_a \!-
\omega_b.  \label{mary4}
\end{equation}

Detailed analysis demonstrates that, in general, the anomalous-cage model
predicts the MARY-dependence on parameters of the spin Hamiltonian similar
to that known in the conventional cage model (conventional diffusion
assisted processes within the well) \cite{Shu5}, however, with replacement
of analytical functions by non-analytical ones.

Most important features of non-analytical MARY-dependence on the parameters
of the spin Hamiltonian, predicted by the anomalous-cage model, can be
demonstrated in the simple limit of relatively weak magnetic interactions: $%
(\delta\omega/w)^{\alpha} \ll l_d, \|\hat l_r\|$. In this limit one can
evaluate MARY with the lowest order of expansion of $Y_r$ [eq. (\ref{mfe3})]
in powers of $\hat H_z^0$:
\begin{equation}
Y_r \approx (l_s/l_0) - (l_s/l_0^2) \mathrm{Tr}[\mathcal{P}_{S}(i \hat
H_{z}^0/w)^{\alpha} \mathcal{P}_{S}],  \label{mfe4a}
\end{equation}
where $l_s = \mathrm{Tr}(\mathcal{P}_s \hat l_r\mathcal{P}_s)$ is the
reactivity in the singlet state and $l_0 = l_s + l_d$. Straightforward
evaluation with the use of eq. (\ref{mfe4a}) gives the expression
\begin{equation}
Y_r = (l_s/l_0) - \mbox{$\frac{1}{4}$}(l_s/l_0^2)\cos
(\pi\alpha/2)|\delta\omega|^{\alpha/2}.  \label{mfe4c}
\end{equation}

The non-analytical dependence $Y_r \sim |\delta\omega|^{\alpha/2}$ is just
the manifestation of anomalous nature of diffusion within the well
(anomalous nature of the cage). Noteworthy is that in the case $\alpha \to 1$
the $\delta\omega$-dependent part vanishes. This is because in eq. (\ref%
{mfe4a}) the spin dependent contribution to $Y_r$ is taken into account in
the lowest order in $H_z^0$ [$\sim (H_z^0)^{\alpha}$)]. At $\alpha = 1$,
however, the term of this order, linear in $H_z$, does not contribute to $%
Y_r $. The non-zero contribution, evidently, results only from the second
order term.

\paragraph{Analysis of RYDMR}

Consideration of the most important specific features of RYDMR can
significantly be simplified in the limit of large $ST_0$-coupling $%
\omega_{ST_0} = \delta\omega \sim \langle A^{\mu} \rangle $ [see eq. (\ref%
{rp1})] and relatively weak microwave field $\omega_1$: $(\omega_{ST_0}/w)^{%
\alpha} \gg \|l_r \|, l_d $ and $(\omega_1/w)^{\alpha} \lesssim \|l_r \|,
l_d $. In this limit quantum coherence effects on evolution of all states
with large splitting ($\sim \omega_{ST_0}$) is negligible \cite{Shu5}, i.e.
their evolution can be treated with balance equations (equations for state
populations).

Coherence effects prove to be important only for four nearly degenerate
pairs of states, of type of the TLS discussed above, which describe
resonances non-overlapping in the considered limit. These four pairs can be
combined into two groups of pairs of these TLSs: $(|\pm\rangle_b
|\mp\rangle_a,\,|T_{\pm}\rangle)$ and $(|\pm\rangle_a
|\mp\rangle_b,\,|T_{\pm}\rangle)$, denoted hereafter as $a_{\pm}$ and $%
b_{\pm}$, respectively. Transitions in TLS-pairs $\mu_{\pm}, \: (\mu = a,
b), $ are associated with those in corresponding separate radical $\mu$.

TLS-states $|\pm\rangle_a |\mp\rangle_b$, corresponding to the zeroth
z-projection of the total spin ($S_z = 0$), are the same for systems $\mu =
a $ and $\mu = b$. However, these systems can be considered as uncoupled
because in the studied limit of large $\omega_{ST_0}$ significantly
efficient transitions in systems $a$ and $b$ occur at different values of $%
\omega$ (i.e. corresponding resonances do not overlap, as it was mentioned
above). For this reason it is possible to distinguish the same states $%
(|\pm\rangle_a |\mp\rangle_b$, belonging to $a$- and $b$-systems, and denote
them as $|a\rangle$ or $|b\rangle$, respectively (the subscript $\pm$ or $%
\mp $ can be omitted as it will be explained below).

The assumed initial population of the singlet state reduces to that of the
above-mentioned states $|a\rangle$ and $|a\rangle$ (in which $S_z = 0$) with
the probability $1/2$. Notice that these states are reactive. The reactivity
matrices $\hat l_r^{\mu}$ are similar for all systems and quite accurately
determined as the two level variant of formula (\ref{mfe4}). These matrices
describe reaction with the same rate approximately equal to $l_s/2$, where $%
l_s = \mathrm{Tr}(\mathcal{P}_s \hat l_r \mathcal{P}_s)$ is the reactivity
in $|S\rangle$-state.

All the TLSs give the same contribution to the total yield $Y_r$, differing
only in the resonance frequency ($\omega_a$ or $\omega_b$) if they
correspond to different radicals. Therefore we can combine the identical
contributions of the TLSs $\mu_{+}$ and $\mu_{-}$ into one $Y_{\mu}$ of two
times larger magnitude and omit subscripts $+$ and $-$ in the notation of
the TLSs and their parameters, as it has been mentioned above. In so doing
we arrive at the representation of $Y_r$ in the form
\begin{equation}
Y_r = Y_a + Y_b  \label{mfe5}
\end{equation}
where $Y_{\mu}, \; (\mu = a, b),$ are given by
\begin{equation}
Y_{\mu} = \mathrm{Tr}\bigg[\mathcal{P}_{\mu} {\hat l_r^{\mu}}\frac{1} {\hat
l_r^{\mu} + l_d + (i\hat H_{\mu}/w)^{\alpha}}\mathcal{P}_{\mu}\bigg]
\label{mfe6}
\end{equation}
with $\mathcal{P}_{\mu} = |\mu \rangle \langle \mu |$ and $\hat l_r^{\mu}
\approx \frac{1}{2}l_s \{\mathcal{P}_{\mu}, \dots \}$.

For simplicity, we also assume that the microwave field $\omega_1$ is weak
enough so that the effects of the $\omega_1$-induced transitions can be
treated perturbatively in the lowest order expansion of $Y_r$ in $\| (\hat
H_{\mu}/w)^{\alpha}/\|\hat l_{r,d}\| \ll 1$.

In these assumptions the yield $Y_{\mu}$ can be evaluated with approximate
expression
\begin{equation}
Y_{\mu} \approx \mbox{$\frac{1}{2}$}(l_s/l_{\mu}) - \mbox{$\frac{1}{2}$}%
(l_s/l_{\mu}^2) \mathrm{Tr} [\mathcal{P}_{\mu}(i \hat H_{\mu}/w)^{\alpha}
\mathcal{P}_{\mu}]  \label{mfe7}
\end{equation}
in which $l_s = \mathrm{Tr}(\mathcal{P}_s \hat l_r\mathcal{P}_s)$ and $%
l_{\mu} \approx \frac{1}{2}l_s + l_d$.

Calculation using eq. (\ref{mfe6}) gives for the magnetic field dependent
part $y_r$, which is called RYDMR spectrum,
\begin{equation}
y_r (\omega) = y_a (\omega - \omega_a) + y_b (\omega - \omega_b),
\label{mfe9}
\end{equation}
where $y_{\mu}, \;(\mu = a, b),$ is written as
\begin{eqnarray}
y_{\mu}(\omega) &=& -\mbox{$\frac{1}{2}$}(l_s/l_{\mu}^2) \mathrm{Tr} [%
\mathcal{P}_{\mu}(i \hat H_{\mu}/w)^{\alpha} \mathcal{P}_{\mu}]
\label{mfe10a} \\
&=& -\frac{\cos (\pi \alpha /2)}{2}\bigg(\frac{l_s}{l_{\mu}}\bigg)^{\!2}\!\!
\frac{(\omega_1/w)^{\alpha}}{(1+ \omega^2/\omega_1^2)^{1-\alpha/2}}.\qquad
\label{mfe10b}
\end{eqnarray}

It is worth noting some important specific features of the RYDMR spectrum
(in the case of anomalous diffusion) predicted by eq. (\ref{mfe10b}):

1) Unlike conventional Markovian migration, anomalous diffusion leads to the
non-analytical dependence of the spectrum on $H_{\mu}$ (i.e. on the
parameters of spin hamiltonian) which can be obtained in the lowest order in
$H_{\mu}$. Naturally, as in the case of MARY, the lowest-order value of
RYDMR vanishes in the limit $\alpha \to 1$ because at $\alpha = 1$,
corresponding to conventional caging, RYDMR amplitude is determined by the
second order term of expansion in $H_{\mu}$.

2) At large $\omega$ (at line wings) the RYDMR resonance contributions $%
y_{\mu} (\omega), \; (\mu = a, b), $ decrease as $y_{\mu} (\omega) \sim
1/\omega^{2-\alpha}$, i.e. slower than the Lorenzian line ($y (\omega) \sim
1/\omega^{2}$).

3) The width of resonances in the spectrum is determined by the amplitude of
microwave field $\omega_1$. In other words these kind of spectra are always
measured in the saturation regime \cite{Abr}.

4) At first sight, the fact that the width is determined by $\omega_1$ is a
consequence of long memory effects on the processes governed by anomalous
diffusion, i.e the absence of the characteristic time in these processes.
Therefore in the presence of such time caused, for example, by the
conventional intraradical spin lattice relaxation, the width seems to depend
on this time. In reality, however, this is not true which can easily be
demonstrated in a simple model assuming this time to result from the spin
independent decay of radicals with the rate $w_0$. In this model the
magnetic field dependent yield contributions $y_{\mu}, \; (\mu = a, b)$ are
still given by eq. (\ref{mfe10a}) but with $(i \hat H_{\mu}/w)^{\alpha}$
replaced by $[(w_0 + i \hat H_{\mu})/w]^{\alpha} - (w_0/w)^{\alpha}$:
\begin{equation}
y_{\mu}(\omega) \sim \frac{1}{\omega_1^2+ \omega^2} \Big[\mathrm{Re}\Big(w_0
+ i \sqrt{\omega^2 + \omega_1^2}\Big)^{\!\alpha}\!\! - w_0^{\alpha}\Big] .
\label{mfe11}
\end{equation}
It is easily seen that in the limit $\omega_1 \gg w_0$ this expression
reproduces eq. (\ref{mfe10b}) while in the opposite limit it predicts $%
y_{\mu}(\omega) \sim 1/(\omega_1^2+ \omega^2)^{1-\alpha}$. Therefore the
presence of the characteristic relaxation time does not lead to the change
of the line width of RYDMR spectra slightly changing only the line shape.

\section{Concluding remarks}

This work concerns the detailed analysis of the specific features of
relaxation kinetics in quantum systems induced by anomalous noise. Two types
of quantum systems are considered, as examples: two-level systems and
radical pairs in a potential well whose recombination is assisted by
anomalous diffusion. The analysis is made with the use of the recently
developed convenient and powerful method based on the CTRWA and the
non-Markovian SLE. It demonstrated some important peculiarities of the
kinetics of the processes under study. First of all, the relaxation kinetics
in both type of systems proved to be strongly non-exponential. More subtle
peculiarities were found in spectral characteristics of these processes: the
line shape, its dependence on the parameters of processes, etc.

In addition, the non-Markovian SLE proved to be very efficient in analyzing
not only simple TLSs but also multilevel quantum systems. As an example of
such systems, recombining RP was considered.

In this work we mainly restricted ourselves to analytical analysis of the
processes, however, as it was point out above, the proposed method also
allows for significant simplification of numerical treatment of the
processes under study, especially in much more complicated multilevel
quantum systems: magnetic clusters, magnetic glasses, etc.

{\acknowledgements The work was partially supported by the Russian
Foundation for Basic Research.}

\bigskip

{\bf Figure captions.}

\bigskip

Fig. 1: The spectrum $I (z) = I(\omega)\omega_0$, where $z = \omega/\omega_0$, calculated in the model (\ref{two3}) [using eq. (\ref{res4c})] for two values of $\xi = \omega_0/(2^{1/\alpha}w)$: $\xi = 0.05$ (a) and $\xi = 0.3$ (b), and different values of $\alpha$: $\alpha = 0.3$ (full), (2) $\alpha = 0.7$ (dashes), and $\alpha = 0.90$ (dots).

\bigskip

Fig. 2: Same as in Fig. 1 but for $\xi = 0.7$ (a) and $\xi = 1.5$ (b).

\bigskip

Fig. 3: The spectrum $I (z) = I(\omega)\omega_0$, where $z = \omega/E_0$, calculated in the model (\ref{two3}) [using eq. (\ref{res4c})] for two values of $\alpha$: $\alpha = 0.5$ (a) and $\alpha = 0.9$ (b), and different values of $r_0 = 2v/\omega_s$: $r_0 = 0.5$ (dots), $r_0 = 1.0$ (dashes), $r_0 = 4.0$ (full) and $r = 8.0$ (dash-dots).

\bigskip

Fig. 4: Population relaxation kinetics $N (\tau)$, where $\tau = E_0 t$, calculated with eq. (\ref{res7}) (full lines) for two values of $r_0 = 2v/\omega_s$: $r_0 = 1$ (a) and $r_0 = 2$ (b), and different values of $\alpha$: (1) $\alpha = 0.95$, (2) $\alpha = 0.90$, (3) $\alpha = 0.85$, and (4) $\alpha = 0.75$. Straight (dashed) lines represent exponential dependence [eq. (\ref{res6})].

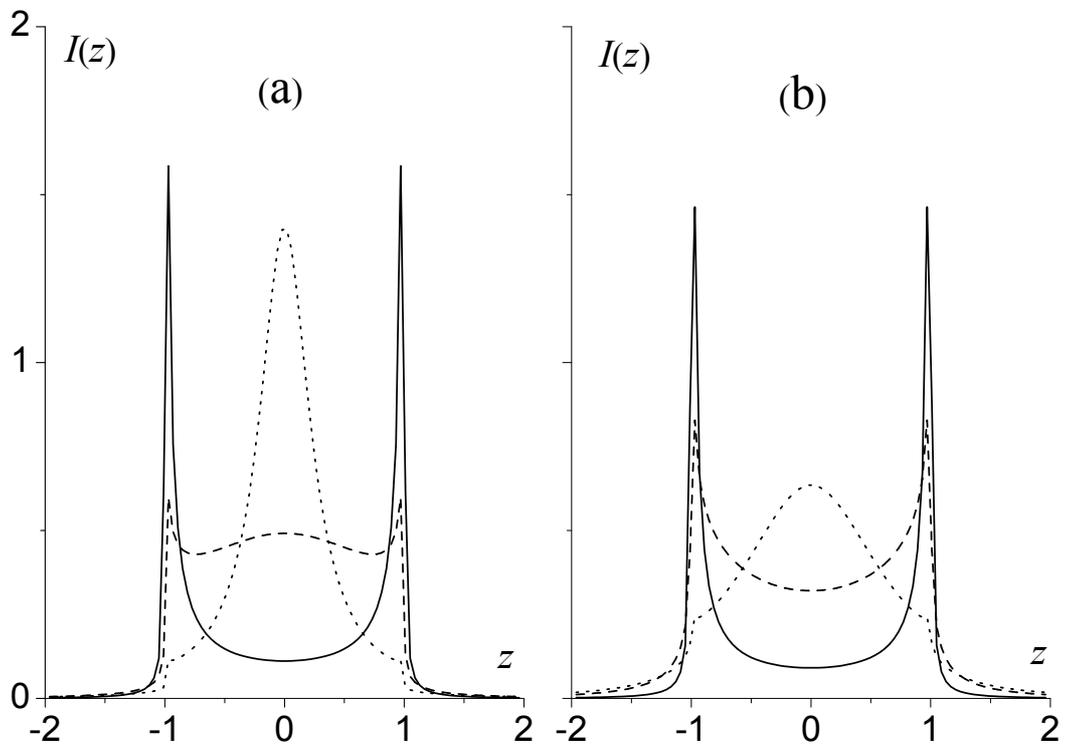
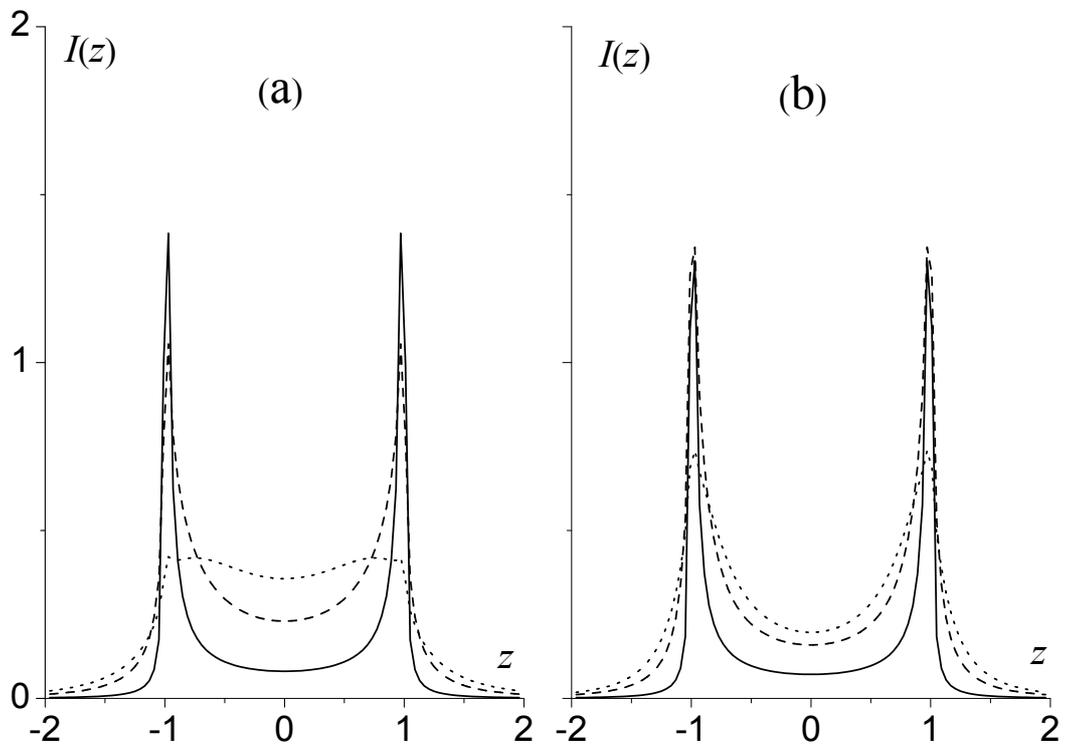
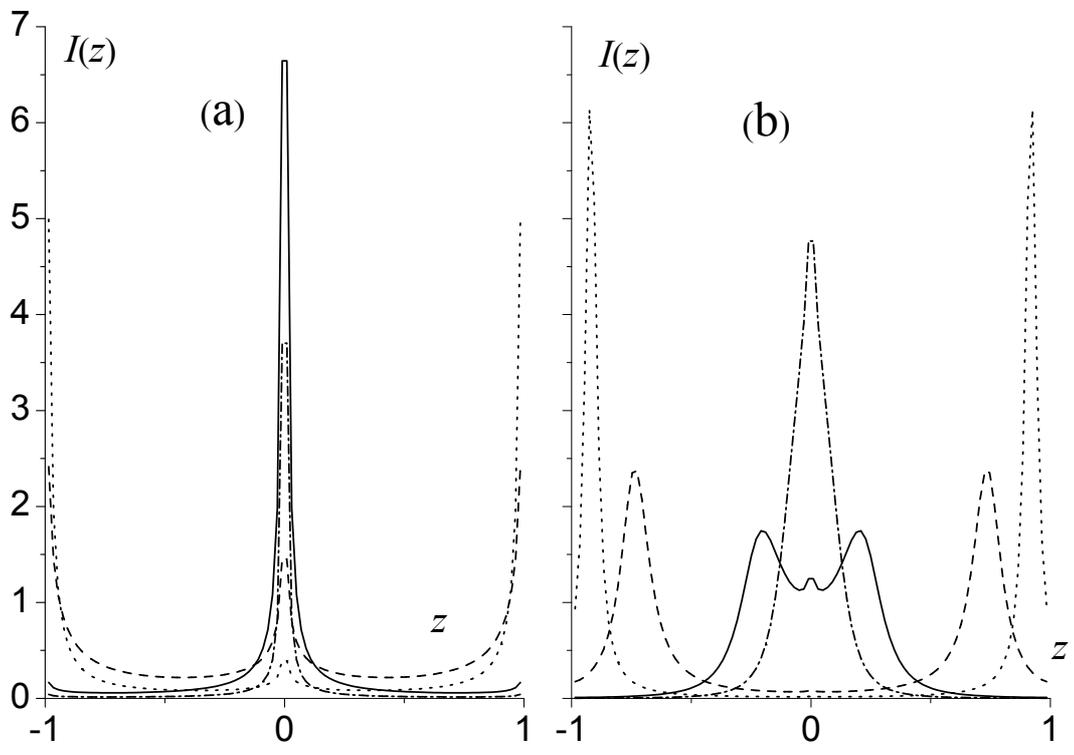
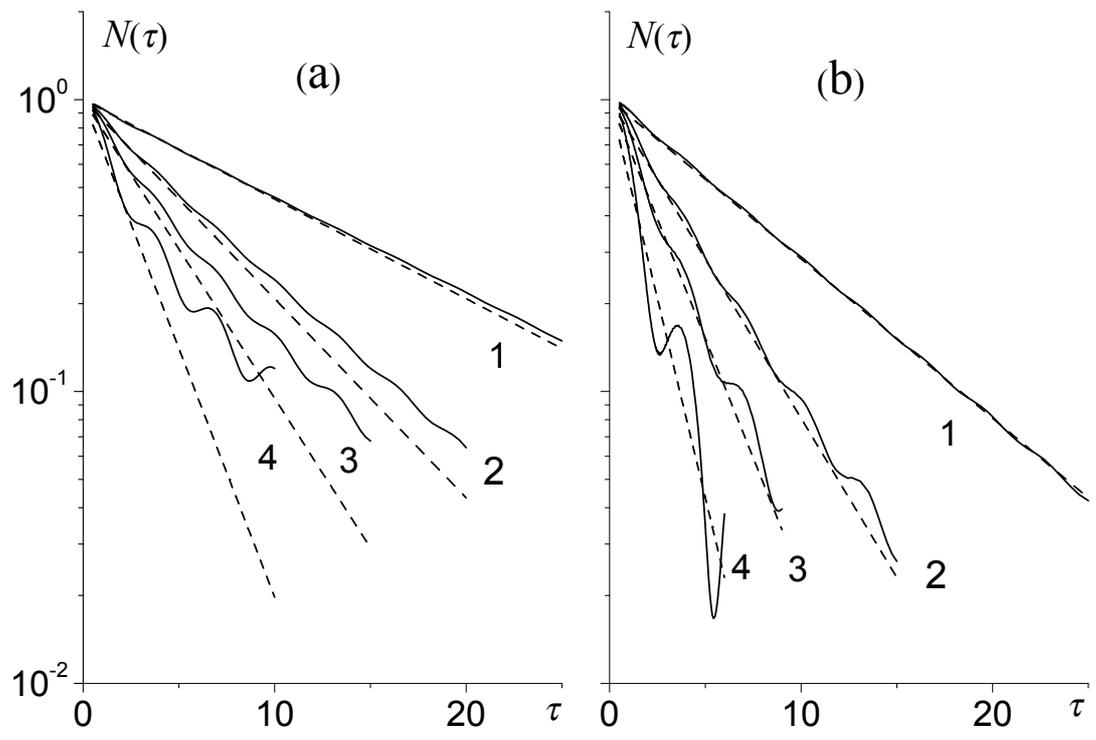
\end{document}